\documentclass[12pt]{article}

\usepackage{amsfonts}
\usepackage{amssymb}
\usepackage{graphics}
\usepackage{epsfig}
\usepackage{multirow}

\usepackage{latexsym}
\usepackage{amsmath}

\usepackage{relsize}
\usepackage{geometry}
\usepackage{color}

\newcommand{\beq}{\begin{equation}}
\newcommand{\eeq}{\end{equation}}
\newcommand{\beqn}{\begin{eqnarray}}
\newcommand{\eeqn}{\end{eqnarray}}
\newcommand{\bea}{\begin{eqnarray}}
\newcommand{\eea}{\end{eqnarray}}
\newcommand{\beas}{\begin{eqnarray*}}
\newcommand{\eeas}{\end{eqnarray*}}

\newcommand{\bquo}{\begin{quote}}
\newcommand{\enqu}{\end{quote}}

\def\Tr{ \hbox{\rm Tr}}

\def\diag{\hbox{\rm diag}}
\newcommand{\gsim}{\lower.7ex\hbox{$\;\stackrel{\textstyle>}{\sim}\;$}}
\newcommand{\lsim}{\lower.7ex\hbox{$\;\stackrel{\textstyle<}{\sim}\;$}}

\def\stroke{\vrule height8pt width0.4pt depth-0.1pt}
\def\topfleck{\vrule height8pt width0.5pt depth-5.9pt}
\def\botfleck{\vrule height2pt width0.5pt depth0.1pt}
\def\Zmath{\vcenter{\hbox{\numbers\rlap{\rlap{Z}\kern 0.8pt\topfleck}\kern
2.2pt
                   \rlap Z\kern 6pt\botfleck\kern 1pt}}}
\def\Qmath{\vcenter{\hbox{\upright\rlap{\rlap{Q}\kern
                   3.8pt\stroke}\phantom{Q}}}}
\def\Nmath{\vcenter{\hbox{\upright\rlap{I}\kern 1.7pt N}}}
\def\Cmath{\vcenter{\hbox{\upright\rlap{\rlap{C}\kern
                   3.8pt\stroke}\phantom{C}}}}
\def\Rmath{\vcenter{\hbox{\upright\rlap{I}\kern 1.7pt R}}}
\def\Z{\ifmmode\Zmath\else$\Zmath$\fi}
\def\Q{\ifmmode\Qmath\else$\Qmath$\fi}
\def\N{\ifmmode\Nmath\else$\Nmath$\fi}
\def\C{\ifmmode\Cmath\else$\Cmath$\fi}
\def\R{\ifmmode\Rmath\else$\Rmath$\fi}

\def\Tr{{\rm Tr}}

\def\diag{{\rm diag}}
\def\2{{1\over 2}}
\def\ntwo{${\cal N}=2\;$}

\def\4N{${\cal N}=4$}
\def\N{{\mathcal N}}

\def\beq{\begin{equation}}
\def\eeq{\end{equation}}
\def\ba{\beq\new\begin{array}{c}}
\def\ea{\end{array}\eeq}

\newcommand{\dm}{|\Delta m |}

\def\Tr{ \hbox{\rm Tr}\,}

\def\SU{{\rm SU}}

\def\R{{\rm R}}
\def\S{{\rm S}}

\def\U{{\rm U}}

\def\diag{\hbox{\rm diag}}

\def\Z{\mathrm Z}
\def\1{\mathbbm{1}}
\def\N{{\cal N}}

\def\C{\rm C}

\def\1{\mbox{\tiny (1) }}
\def\0{\mbox{\tiny (0) }}

\begin{document}

\begin{titlepage}

\begin{flushright}
FTPI-MINN-09/42,  UMN-TH-2825/09 , DAMTP-2009-82 \\
January 12/2010

\end{flushright}

\vspace{1mm}

\begin{center}
{\large  {\bf Higher Winding Strings and \\[1mm] Confined Monopoles in 
\boldmath{$\mathcal{N}=2$} SQCD }}
\end{center}

\vspace{0.1mm}

\begin{center}
{\large  R. {\sc Auzzi}$^{(1)}$}, {\large S. {\sc Bolognesi}$^{(2)}$} and
{\large M. {\sc Shifman}
$^{(3)}$}  

\vspace{0.1mm}

$^{(1)}${\it \footnotesize
Racah Institute of Physics, The Hebrew University,  \\ Jerusalem 91904, Israel}
\\[1mm]
$^{(2)}$ {\it \footnotesize  DAMTP, 
Center for Mathematical Sciences, \\
Wilberforce Road, 
Cambridge, CB3OWA, UK }
 \\[1mm]  $^{(3)}${\it \footnotesize
William I. Fine Theoretical Physics Institute, University of Minnesota, \\
116 Church St. S.E., Minneapolis, MN 55455, USA}
\end{center}

\begin{abstract}

We consider composite string solutions in $\N=2$ SQCD with the gauge group U$(N)$,
the Fayet--Iliopoulos term $\xi \neq 0$ and $N$ (s)quark flavors. These bulk theories support non-Abelian strings
and confined monopoles identified  with kinks in the two-dimensional world-sheet theory.
Similar and more complicated kinks (corresponding to composite confined monopoles)
must exist in the world-sheet theories on composite strings. In a bid to detect them we 
analyze the Hanany--Tong (HT) model, focusing on a particular example of $N=2$. 
Unequal quark mass terms in the bulk theory result in the twisted masses in the 
$\N=(2,2)$ HT model. For spatially coinciding 2-strings, we
 find three distinct minima of potential energy,
corresponding to three different 2-strings.
Then we find BPS-saturated kinks interpolating 
between each pair of vacua. Two kinks can be called elementary. They emanate one unit of the magnetic flux
and  have the same mass
as the conventional 't Hooft--Polyakov monopole on the Coulomb branch of the bulk theory ($\xi =0$).
The third kink represents a composite bimonopole, with
twice the minimal magnetic flux. Its mass is twice the mass of the elementary confined monopole.
We find instantons in the HT model, and discuss quantum effects in composite strings at strong coupling.
In addition, we study the renormalization group flow in this model.

\end{abstract}

\end{titlepage}

\vfill\eject 

\section{Introduction}

Non-Abelian strings in a class of 
four-dimensional \ntwo\, gauge theories were discovered and explored  recently 
\cite{ht1,ABEKY,SYmon,ht2} (for reviews see  \cite{Trev}). 
In addition to translational (and supertranslational) moduli
characterizing the position of the string center in the perpendicular plane,
non-Abelian strings are endowed with orientational (and superorientational)  moduli on the string world sheet.
The orientational moduli emerge from the fact that the bulk theories 
supporting such strings possess a color-flavor locked SU$(N)_{c+f}$ 
global symmetry
while a particular string solution preserves only an  $\SU(N-1) \times \U(1)$ subgroup.
 Therefore, in fact, we deal with a 
 $\mathbb{CP}^{N-1}$
 family of solutions; the orientational moduli describe how each particular string solution from this family
is embedded in SU$(N)_{c+f}$. These strings are BPS saturated,
 and the worldsheet theory retains $\N=(2,2)$ supersymmetry. 
As a result, holomorphy
protects certain (chiral) quantities, such as tensions, 
which  are then exactly calculable.

Soon after the non-Abelian strings, it was discovered that kinks in the world-sheet 
theories on non-Abelian strings  describe confined monopoles \cite{SYmon,ht2}. 
These kinks cannot detach themselves from the strings and can be at strong coupling   
even in the weakly coupled bulk theory.  
This observation provides a physical, and very transparent, explanation 
for the earlier detected coincidence of the BPS spectra of two theories \cite{Dorey:1998yh}:
the one on the world sheet and the four-dimensional \ntwo\, theory 
in the $r=N$ vacuum 
on the Coulomb branch.

Deformations of various parameters of the bulk theory present an excellent 
research laboratory.
The gauge symmetry of the bulk \ntwo theories is U$(N)$,
and  they have $N$ quark flavors
(i.e. $N$ hypermultiplets in the fundamental representation).
Moreover, they are endowed with the Fayet--Iliopoulos (FI) term $\xi$. If
$\xi\gg \Lambda^2$ 
the bulk theory is at weak coupling (here $\Lambda$ is the scale parameter of \ntwo SQCD).
Other dimensional parameters of the bulk theory are the (s)quark mass
terms. Physically observable are the differences
$\Delta m = m_i-m_j$. 
As was mentioned,  the world sheet theory is \cite{SYmon,ht2}  
 $\mathbb{CP}^{N-1}$ sigma model.
In fact, if $\Delta m \neq 0$, we deal with the  $\mathbb{CP}^{N-1}$ model
with twisted masses \cite{twisted}.

One can start from $\xi=0$ and $|\Delta m |$ large
(compared to $\Lambda$), and continuously deform
$\xi$, increasing its value, and, simultaneously,  decreasing $\dm$.
One can trace this deformation from the beginning to the end.
At $\xi=0$ we have conventional 't~Hooft--Polyakov monopoles,
then, as $\xi$ increases, the non-Abelian strings are formed and attach themselves to the 
't~Hooft--Polyakov  monopoles squeezing their magnetic flux into flux tubes.
The tension of the flux tubes grows and they become thinner while the monopoles
become exceedingly fuzzier albeit they retain their BPS nature. At the end, at $\sqrt\xi \gg \dm$,
they turn into kinks
in the world-sheet theory. The mass of the 
monopoles/kinks does not depend on $\xi$. 
At $\dm \gg \Lambda$
this mass stays the same independently of whether  the monopoles
are confined or unconfined. The deformation process is described in detail in
\cite{SYrev}.

 Let us discuss in more detail the bulk theory which has the  U(2) gauge group and two flavors. 
 If $\Lambda \ll \dm \ll \sqrt{\xi}$, 
quantum fluctuations on the string world sheet are tempered, and
two distinct
elementary strings (i.e. those with the minimal tension
$2\pi\xi$)  are easily identifiable.
The SU(2) orientational moduli (described by O(3)=$\mathbb{CP}^{1}$ model with the twisted masses)
weakly fluctuate around two (vacuum) points: either
$S_3=1$ or  $S_3=-1$, i.e. the flux is oriented in the group space in the direction
of either the north or south pole.\footnote{   We will refer to them as $| \pm \rangle$ states.
Needless to say,
geometrically both magnetic fields, from  $\U(1)_0$ and $ \U(1)_3$, are aligned along the string axis.}

\begin{figure}[h!t]
\epsfxsize=8cm
\centerline{\epsfbox{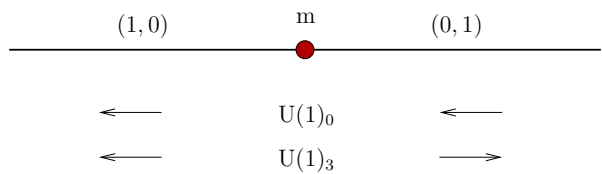}}
\caption{{\footnotesize The confined monopole is a kink that changes the string state
from $| + \rangle$ to $| - \rangle$ or vice versa. }}
\label{fone}
\end{figure}

The magnetic flux has the following decomposition in terms of $\U(1)_0$ and $\U(1)_3$:
\bea
(1,0):  &&\quad
\frac{1}{2}\left(
\begin{array}{ccc}
 1 &   \\
     & 1     
\end{array}
\right)_0  + \frac{1}{2} \left(
\begin{array}{ccc}
 1 &     \\
   &  -1    
\end{array}
\right)_3 = \left(
\begin{array}{ccc}
1  &      \\
  & 0     
\end{array}
\right)\ ;\nonumber \\ [3mm]
(0,1):  &&\quad \frac{1}{2} 
\left(
\begin{array}{ccc}
 1 &   \\
     & 1     
\end{array}
\right)_0  - \frac{1}{2} \left(
\begin{array}{ccc}
 1 &     \\
   &  -1    
\end{array}
\right)_3 = \left(
\begin{array}{ccc}
0  &      \\
  & 1     
\end{array}
\right) \,,
\label{dom2}
 \eea
 where the subscript 3 marks the U(1) subgroup generated by the third generator of SU(2).
We call these strings $(1,0)$ and $(0,1)$, respectively,
since in the former case it is only the first flavor that winds, while in the latter case it is the
second flavor. 
Note that a ``basic" winding in $\U(1)_0$
for the non-Abelian string is by $\pi$ rather than by the conventional
$2\pi$ of the Abrikosov--Nielsen--Olesen (ANO) string.  
But the sum of the two $\U(1)$ windings (in $\U(1)_0$ and $\U(1)_3$)
 creates an ordinary  $2\pi$ winding locked to the first flavor or to the second.
If the ${\rm U}(1)_0$ magnetic field $\vec B$ inside the string points from right to left,
then in the $(1,0)$ string
the $\U(1)_3$ magnetic fields $\vec B^{\, 3}$ is directed  from right to left too
while it is directed from left to right  in the $(0,1)$ string.
The combined $\vec B^{\, 3}$ magnetic flux for
two strings attached to the kink (which either inflows or outflows the kink,
depending on whether we have the $(1,0)$-$(0,1)$ or $(0,1)$-$(1,0)$ string
junction) is one unit of the magnetic monopole flux. 
The monopole carries flux under $\U(1)_3$. This is depicted in Fig.~\ref{fone}.

\begin{figure}[h!t]
\epsfxsize=11cm
\centerline{\epsfbox{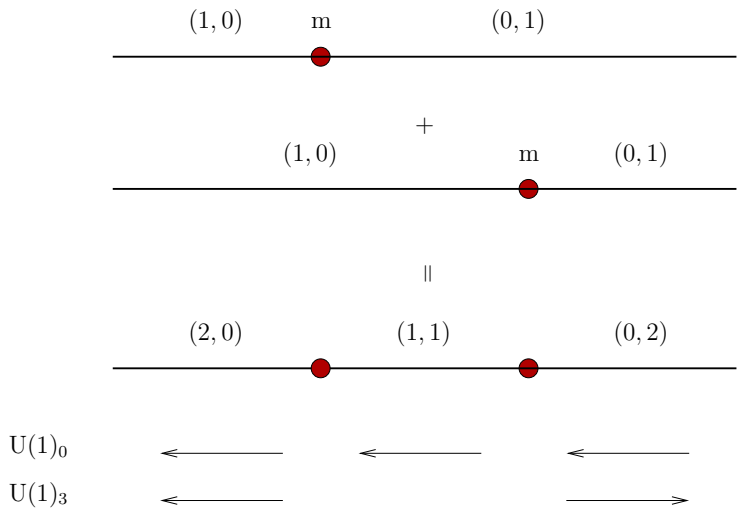}}
\caption{{\footnotesize Two monopoles can be confined on a composite string as a composite kink. }}
\label{ftwo}
\end{figure}

Given the confined-monopole/kink correspondence outlined above,
it se\-ems necessary and timely to address two questions:
(a) manifestation of the unit-flux monopoles in {\em composite} strings; 
(b)  {\em multiple monopole} configurations. 
We will show that mono\-poles with the {\em unit} magnetic charge
manifest themselves  as  junctions of the type (2,0)-(1,1), while
{\em multi}monopole states, with the magnetic charge 2 and higher, exist as 
a chain of junctions of the {\em composite} strings.
It is impossible to confine two monopoles\,\footnote{We mean here 
two monopoles rather than
the monopole-antimonopole pair with the vanishing net magnetic charge.}
on the elementary
non-Abelian string. Magnetic charge-2 configurations
 necessarily belong to  composite strings built of  two (or more) 
constituent strings. We explicitly construct,
in the U(2) bulk theory with 
two coaxial elementary  strings, 
 a continuous family of composite kink solutions (2,0)-(1,1)-(0,2). This is depicted in Fig.~\ref{ftwo}.

\vspace{5mm}

If $\dm = 0$,
the two-string configuration acquires a compact part of the moduli space
associated with the relative orientations in the group space. Switching
$\Delta m \neq  0$ we lift the continuous degeneracy
of this part of the moduli space.
One of the goals of this paper  is to trace how exactly the moduli space 
of multiple strings is affected by the twisted mass deformation.

Assume we have two separate elementary 
strings (at rest) at a certain fixed distance  distance $L$ from each other.
How many states  this system has?  
Since all bulk excitations are massive (there is a mass gap in the bulk)
the thickness $\ell$  of each elementary string is finite and is related
to the inverse masses of the bulk particles. We assume that $L>\ell$.
Since each can be in two different states we have a total of four states. 
The four states can then be grouped in  three possible two-string configurations,
\beqn
{\rm (i)}\,\,\qquad\qquad && \quad \; (1,0) +(1,0)\ ;  
\label{B2}\\[1mm]
{\rm (ii)}\,\qquad\qquad && \quad \; (0,1) +(0,1)\ ; 
\label{B3} \\[1mm]
{\rm (iii)}\qquad\qquad &&
\left\{
\begin{array}{c}
  (1,0) +(0,1)\ ; \\[1mm]
  (0,1) +(1,0) \ .
\end{array}
\right.
\label{Bfour}
\eeqn
In all three cases, if we take a large circle encompassing both strings in the perpendicular
plane, the U(1)$_0$ winding of the matter fields is $2\pi$. This winding is noncontractible.
In the first two cases   (\ref{B2}),  (\ref{B3}) the $\U(1)_3$-winding in SU(2)  is $\pm 2\pi$.
It is topologically contractible to no winding in SU(2). (There is a potential barrier, however, 
determined by $\Delta m\neq 0$.) In the third case the overall $\U(1)_3$-winding in
SU(2) can be contracted to no winding
without any barrier. The ANO string is a part of this sector, with no separating barrier.
The configurations are dynamically stable.
A way to see that the last two must belong to the same sector, is to realize that they 
can be connected by a physical exchange of two strings. 

If the two-string configuration above are BPS saturated,\footnote{At $L\to\infty$
all three configurations, (i), (ii) and (iii) above, are BPS saturated. 
 Since the multiplet is short in \ntwo,
the property of the BPS saturation cannot disappear as we vary $L$.}
the tensions of the composite objects is $4\pi\xi$, i.e. twice the tension
of the elementary strings.

\begin{figure}[h!t]
\epsfxsize=13cm
\centerline{\epsfbox{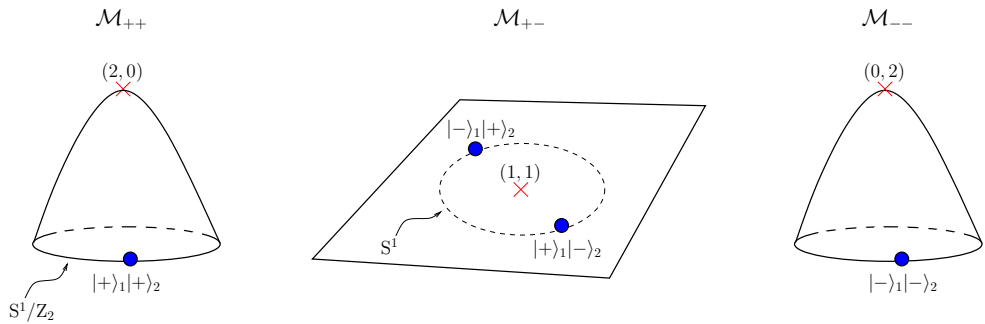}}
\caption{{\footnotesize The moduli space of vortices for the mass deformed 
theory, for $n=2$, has three disconnected components: 
${\cal M}_{++}$, ${\cal M}_{+-}$, and ${\cal M}_{--}$. }}
\label{4vs3}
\end{figure}

The elementary string has two ground states, $| \pm \rangle$.  
Since each of  two strings can be in two different states we have a total of four states. 
The moduli space (at $\Delta m \neq 0$) has only three disconnected components, not four (Fig.~\ref{4vs3}).
Two states (\ref{Bfour}) belong to one and the same manifold  ${\mathcal M}_{+-}$. 
They could be classified according to interchange symmetry. However, when the
inter-string distance $L$ tends to zero, only one state survives on ${\mathcal M}_{+-}$.
Therefore, in our set up, we will deal with three 
distinct  composite strings corresponding to three points marked by ``x"
in three plots in Fig.~\ref{4vs3}. The manifolds ${\mathcal M}_{++}$ and ${\mathcal M}_{--}$ 
are similar to the moduli space of double vortex in 
the U$(1)$ theory \cite{Samols:1991ne}. Asymptotically it
is the cone obtained from the complex plane modulus by a $Z_2$ reflection. 
The singularity at the tip of the cone is resolved at the scale of the string thickness. 
This  implies, in particular,  the $\pi/2$ scattering for head-on-head collisions. 

The manifold ${\mathcal M}_{+-}$ does not have this $Z_2$ factorization 
and presents   a plane asymptotically. In the head-on-head scattering the two 
strings pass one trough the other,
 and the scattering angle is $\pi$ rather than $\pi/2$.

Solutions for the solitonic  2-strings with the coinciding axes in the given bulk 
theory were found and studied previously
\cite{knp,knp2,mmatrix,knp3} for $\Delta m =0$. The reduced moduli 
space (with $L=0$) was shown \cite{knp2,knp3} to be topologically equivalent to
$\mathbb{CP}^{2}/{Z}_{2}$. The metric of the full moduli space, including 
the collective coordinates associated with 
$L\neq 0$, remains unknown. Unlike the metric for the elementary string moduli space,
for composite strings it cannot be determined on the basis of symmetry considerations
due to entanglement of the orientational and translational moduli.
What is available at the moment
is a model suggested by Hanany and Tong \cite{ht1,ht2} who
embedded the bulk gauge theory in a stringy set-up made of intersections of 
D4 and NS5 branes in type IIA string theory.
The bulk gauge theory of interest is defined as a certain decoupling limit of the 
low-energy description of the D4 branes.
The flux tubes then correspond to D2 branes. The $(1+1)$-dimensional world-sheet theory 
 is  a $\U(k)$ sigma model with $\N=(2,2)$,  one adjoint field  $Z$ and $N$ fundamentals 
 $n$.\footnote{We will consider the case $N=k=2$.}

The Hanany--Tong (HT) model admittedly captures only some features of the 2-string 
solutions. For instance,
at large $L$ the string interaction in the HT model falls off in a power-like manner, while
in fact, with the gapped bulk theory, it should fall off exponentially.
It was argued, however, that the HT model is in the same universality 
class as the  (unknown) genuine world-sheet theory and, therefore, 
 correctly describes holomorphic quantities  and
reproduces  physics of the BPS objects.
We will use the HT model (with the twisted masses switched on)
 just for these purposes.
Our findings can be seen as a confirmation that it works well in this context.

\vspace{1mm}

Our main results can be summarized as follows. We introduce the twisted masses in the 
$\N=(2,2)$ HT model,
and find three distinct minima of the potential energy,
corresponding to three different 2-strings (i) -- (iii).
Acting in the subspace $L=0$ of the moduli space we find BPS-saturated kinks interpolating 
between each pair of vacua. Two kinks interpolating between (2,0) and (1,1) and
(1,1) and (0,2) can be called elementary. They emanate one unit of the magnetic flux.
In essence, they are the same confined monopoles
as those found in \cite{SYmon,ht2}. They have the same mass as the kinks in \cite{SYmon,ht2},
which, in turn, have the same mass
as the conventional 't~Hooft--Polyakov monopole 
in the $r=N$ vacuum 
on the Coulomb branch of the bulk theory ($\xi =0$).
The kink interpolating between (2,0) and (0,2) represents a composite monopole, with
twice the minimal magnetic flux. Its mass is twice the mass of the elementary confined monopole
(see the bottom part of Fig.~\ref{ftwo}.)

We discuss instantons  effects
in composite strings in the limit $\Delta m \to 0$.
We are able to find explicit instanton solution  in the Hanany--Tong model.
 At $L\to 0$, this is the strong coupling limit on the 
world sheet. We argue
that the quantum moduli space of two coincident strings  is in fact built of
three disconnected components.
 
 Finally, we study the renormalization group flow.
 
 \vspace{2mm}
 
The paper is organized as follows.
 In Sect.~\ref{theosetting} we briefly review  the basic  bulk theory supporting non-Abelian strings.
 We review both, elementary strings and what is known about composite strings of nonminimal winding.
 In Sect.~\ref{httheo} we introduce the Hanany--Tong model including the twisted mass deformation. 
 The limits of validity of the HT model following from the string set up
 are discussed. We then explore in detail the moduli space of composite vortices, with 
 the twisted-mass-generated potential, at $L=0$. Three isolated supersymmetric vacua are identified.
 Section~\ref{spoex} treats the spectrum of excitations. There are elementary excitations -- oscillations near the vacua.
 Of more interest to us are solitonic excitations -- BPS kinks -- on which we focus. 
In Sect.~\ref{costr} we discuss the
limit $\Delta m \ll \Lambda$ in which dynamics is determined by strong quantum effects.
Section \ref{rg} is devoted to  quantum effects  from the standpoint   of the sigma-model renormalization-group flow.
Section~\ref{conclu} summarizes our findings. In Appendix we consider strings with the opposite directions of 
$\vec{B}^3$ and generic $L$ (i.e. $L\neq 0$).

\section{Flux tube in four dimensions}
\label{theosetting}

\subsection{Theoretical setting}
\label{wthse}

We consider $\mathcal{N}=2$ SQCD with $N_f=N_c=N=2$ in the bulk,
with the  Fayet--Iliopolous term ($D$ term)
and masses for the quark
hypermultiplets,
\beq
 m_1=-m_2=m \, . 
 \label{twisma}
 \eeq
 The original gauge group is U(2).
The bosonic part of the action (in the Euclidean notation) is
\beqn
{\cal L} &=& \int d^4  x \bigg[ \frac{1}{4 e_3^2} |F_{\mu \nu}^k|^2 +\frac{1}{4 e_0^2} |F_{\mu \nu}|^2+
 \frac{1}{ e_3^2} |D_{\mu} a^k|^2+ \frac{1}{ e_0^2}  |\partial_{\mu} a|^2
 \nonumber\\[3mm]
 &+& \Tr (\nabla_\mu Q)^{\dagger} (\nabla_\mu Q)+
 \Tr (\nabla_\mu \tilde{Q}) (\nabla_\mu \tilde{Q}^{\dagger})+
 V(Q,\tilde{Q},a^k,a) \bigg],
 \label{azione-tutta}
 \eeqn
 where $e_0$ and $e_3$ are the gauge couplings for U(1) and SU(2) factors, respectively, and 
\beqn
V&=& \frac{e_3^2}{8}
 \left( \frac{2}{e_3^2} \epsilon^{ijk} \bar{a}^{j} a^k +
 \Tr (Q^\dagger \sigma^i Q) -\Tr(\tilde{Q} \sigma^i \tilde{Q}^\dagger) \right)^2
 \nonumber\\[3mm]
  & +&
 \frac{e_0^2}{8} \left(\Tr (Q^\dagger Q)-\Tr(\tilde{Q} \tilde{Q}^\dagger)- 2  \xi \right)^2
 \nonumber\\[3mm]
  & +& \frac{e_3^2}{2} \left|\Tr (\tilde{Q} \sigma^i Q)\right|^2 +
 \frac{e_0^2}{2} \left|\Tr (\tilde{Q}  Q )    \right|^2
 \nonumber\\[3mm]
  & +&  \frac{1}{2} \sum_{f=1}^2 |(a+\sigma^i a^i- m_f) Q_f |^2+
 |(a+\sigma^i a^i-m_f) \tilde{Q}^{\dagger}_{f} |^2 \,.
 \eeqn
The vacuum expectation values (VEVs) 
of the squark fields are given by the following expression:
\beq
Q=\sqrt{\xi} \left(\begin{array}{cc}
1 & 0\\
0 & 1 \\
\end{array}\right) \, , \qquad \tilde{Q}=
\left(\begin{array}{cc}
0 & 0\\
0 & 0 \\
\end{array}\right) \,
, \qquad a_3=m \, .
\eeq
For a thorough review see \cite{SYrev}.

\subsection{Minimal-winding flux tube}
\label{dom3}

The minimal-winding vortex solution c an be found using the {\em ansatz}
\beqn
  Q
  &=&
  \left(\begin{array}{cc}
\phi_1  e^{i \varphi}& 0\\[2mm]
0 & \phi_2 \\
\end{array}\right) \, , 
\nonumber\\[3mm]
 {A_i}
 &=&
  \frac {  \epsilon_{ij}    x_j} {r^2}   \left(    \sigma^3   \frac{1-f_3}{2}+ 1  \frac{1-f}{2}   \right) \, .
 \label{wansatz}
\eeqn
The classical solution is $1/2$ BPS-saturated leaving four supercharges unbroken. 
Using a color+flavor rotation, we can write a family of solutions,
\beqn
  Q
  &=&
  U\cdot \left(\begin{array}{cc}
\phi_1  e^{i \varphi}& 0\\
0 & \phi_2 \\
\end{array}\right) \cdot U^\dagger=\frac{\phi_1 e^{i \varphi}+\phi_2}{2} 1 + n^a  \sigma^a \frac{\phi_1  e^{i \varphi} -\phi_2}{2} \, ,
\nonumber\\[3mm]
  {A_i}
  &=&
   \frac {  \epsilon_{ij}    x_j} {r^2}   \left[  \left(n^a  \sigma^a\right)   \frac{1-f_3}{2}+ 1  \frac{1-f}{2}   \right] , 
\label{wcloso}
\eeqn
where $U$ is an arbitrary SU(2) matrix, and $n^a$ parametrize the internal
$\mathbb{CP}^1$ moduli.
Moreover, $x_j$ ($j=1,2$) parametrizes two coordinates in the perpendicular plane.
For general $N$, the compact part of the classical  moduli space is obviously 
\beq 
\frac{{\rm SU}(N)_{c+f}}{\left({\rm U}(1) \times {\rm SU}(N-1)\right)_{c+f}}=\mathbb{CP}^{N-1} \, ,
\label{copams}
\eeq
rather than $\mathbb{CP}^{1}$.
Next, we promote the  classical moduli   to   fields living on the string world sheet.
The resulting effective theory is the $\N=(2,2)$ $\mathbb{CP}^{N-1}$ sigma model.
The quark mass terms (more exactly, their differences)
descend to the world sheet
in the form of the twisted masses. 

\subsection{Two coincident strings}

Using the index theorem, one can  show \cite{ht1} that
in the $\mathcal{N}=2$ theory with $N_c=N_f=N$,
the moduli space of the winding-$k$ vortices 
is a manifold with real dimension
$ 2 k N $. In the limit of large distance between the $k$ elementary vortices,
this has a simple interpretation: $2 k$ of these coordinates correspond to the position
of each elementary string (translational moduli)
 while  $2 k (N-1)$ correspond to the orientation of each constituent
in the internal $\mathbb{CP}^{N-1}$ space (orientational moduli).

As was mentioned, we focus  on the case  $k=N=2$.
An explicit solution for two coincident vortices was found in \cite{knp2} by virtue of
the {\em ansatz}
 \beqn
 Q 
 &=&
   \left(\begin{array}{cc}
-\cos \frac{\gamma}{2}  e^{ 2 i \varphi}  \kappa_1
& \sin \frac{\gamma}{2}  e^{  i \varphi} \kappa_2  \\[3mm]
- \sin \frac{\gamma}{2}  e^{  i \varphi}  \kappa_3
& -\cos \frac{\gamma}{2} \kappa_4  \\
\end{array}\right),
 \label{ans1}
\\[4mm]
A^0_{(i)} 
&=&
 -\frac{\epsilon_{ij} x_j}{r^2}  (2-f_0)\,, \qquad
 A^3_{(i)} = -\frac{\epsilon_{ij} x_j}{r^2}  \left[(1+\cos \gamma )-f_3\right],
 \label{ans2}
\\[3mm]
A^1_{(i)} 
&=& 
-\frac{\epsilon_{ij} x_j}{r^2} (\sin \gamma) (\cos\varphi)  (1-g), 
 \label{ans3}
 \\[3mm]
A^2_{(i)} 
&=&
 +\frac{\epsilon_{ij} x_j}{r^2} (\sin \gamma) (\sin \varphi)  (1-g)\,,
 \label{ans4}
 \eeqn
 where the functions $\kappa_{1,2,3,4}$, $f_{0,3}$, and $g$
 depend only on $r=\sqrt{x_1^2+x_2^2}$, the angle $\varphi$ is the polar angle in the
 plane  perpendicular to the string axis, while $\gamma$ is the angle characterizing the
 relative group orientation of two strings comprising the 2-string in question
 (for further details see \cite{knp2}).
Now we can apply an ${\rm SU}(2)_{c+f}$ rotation to this solution.
For generic $\gamma$ all   generators of this symmetry are spontaneously
broken on the string. 
Thus, the moduli space for coincident strings has  dimension four.
A more general solution, corresponding to strings with
arbitrary orientation and relative separation,
can be found in the framework of the moduli matrix approach \cite{mmatrix}.

It is difficult to carry out an honest-to-god derivation
of the 
 $\N=(2,2)$ sigma model on the 2-string world sheet directly from the bulk theory.
The world-sheet description involves a sigma model with a highly
non-trivial metric,  not determined by the symmetries of the problem.
With  nonvanishing masses for the quark hypermultiplets,
(with $\dm\ll\xi$), in addition,  there is  a nontrivial potential
on the moduli space, which is also difficult to calculate in full
from the four-dimensional theory. 

In the absence of a genuine world-sheet
model derived from the first principles
we will settle for a simplified substitute believed to   describe well
some crucial aspect of the world-sheet physics. 

\section{Two-dimensional effective theory}
\label{httheo}

\subsection{The brane construction}

\begin{figure}[h!t]
\epsfxsize=8cm
\centerline{\epsfbox{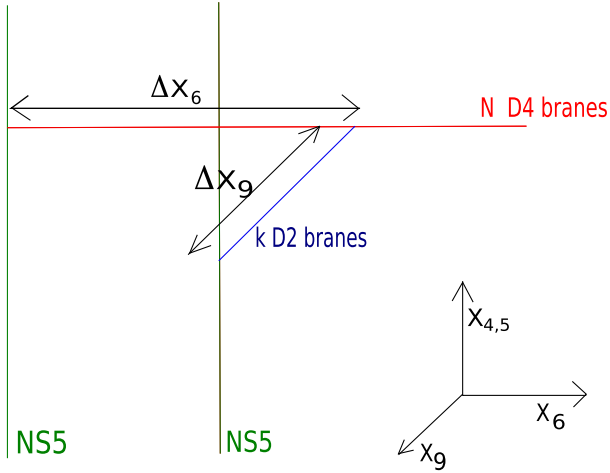}}
\caption{{\footnotesize The brane set-up in Type IIA string theory.
The $k$-string configurations correspond to $k$ D2-branes stretching between 
the NS$5$ and  $N$ D$4$-branes.
$N$  chiral multiplets $n_j$ in the fundamental representation
arise from fundamental strings stretching between the D$4$ and the D$2$ branes.}}
\label{fonep}
\end{figure}

To begin with, let us
 briefly review  the Hanany--Tong construction \cite{ht1,ht2},
based on the string theory/brane realization \cite{hw,witten97}, type IIB or A, for   $2+1$ or $3+1$-dimensional bulk, respectively. 
Focusing on the latter case, we start from
two parallel NS$5$-branes extended in
the  directions $x^{0,1,2,3,4,5}$ and 
separated  by some distance $\Delta x^6$ in the direction $x^{6}$ (see Fig.~\ref{fonep}).   The gauge  D$4$-branes 
(we have $N$ such branes)
are extended in the directions $x^{0,1,2,3}$ and $x^6$,
 between the above two NS$5$-branes. Moreover,
 $$
 1/e^2 \sim \frac{\Delta x^6}{g_s l_s}\,,
 $$ 
 where $e$ is the induced gauge coupling, and the flavor D$4$-branes are 
 semi-infinite in $x^{6}$ and attached only to one of the
NS$5$-branes, say NS$5'$. 
When the gauge and flavor branes are locked, the NS$5'$ can be moved;
 a global translation in the $x^9$ direction corresponds to the induced FI term 
 $$\xi \sim \frac{\Delta x^9}{ g_s l_s^3}\,.$$
The field theory living on  $x^{0,1,2,3}$ of the D$4$-branes, is obtained by the decoupling of the 
Kaluza--Klein modes ($\sim 1/\Delta_6$) as well as the string modes ($\sim 1/l_s$).  This 
decoupling limit is 
\beq
\Delta x^6 = \delta_6 g
_s l_s\,, \qquad \Delta x^9 = \delta_9 g_s  l_s\,,  \qquad g_s \to 0\,.
\label{wachie}
\eeq
The scaling formula (\ref{wachie}) reproduces, at energy scales much lower than $1/l_s$, a $3+1$-dimensional
 theory with the fixed values of $e$ and $\xi$.
To be able to consistently include  Higgsing of the bulk theory  we must require 
\beq
\delta_9 \ll 1\,, \,\delta_6\,.
\label{tbad}
\eeq
To impose  the classical limit $e \to 0$, it is necessary to have  
\beq
\delta_6  \gg 1\,.
\label{tbadd}
\eeq
We do not take into account strong coupling effects here.

In this set-up the flux tubes correspond to D$2$-branes extended in 
the directions $x^{0,3,9}$ and stretched between  one NS$5$-brane and $N$ 
distinct D$4$ branes.
As a result, 
the two-dimensional $\N=(2,2)$ theory 
on the world sheet of $k$ parallel strings is a $\U(k)$ gauge theory with one chiral multiplet $Z$ in the adjoint representation
 (which corresponds to the position moduli of the vortex strings on the transverse plane $x_{1,2}$) 
 and $N$  chiral multiplets $n_j$ in the fundamental representation
(which arise from fundamental strings stretching between the D$4$ and the D$2$ branes).
The adjoint gauge multiplet is the remnant of the D$2$-brane gauge theory, compactified on the segment $\Delta x^9$.
The flavor multiplet corresponds to the strings with one end on the D$2$-branes and the other on the D$4$-branes.
For the strings far apart, 
$$
\U(k) \to \U(1)^k\,,
$$ 
and 
$$Z=\diag(Z_1,\dots, Z_n)\,.$$
The theory  reduces to $k$ distinct factorized
$\N=(2,2)$ $\mathbb{CP}^{N-1}$
 models.
The induced gauge coupling and 
the FI term of this two-dimensional
theory are 
$$1/g^2 \sim \frac{\Delta x^9 l_s }{ g_s}\, ,\qquad r \sim \frac{\Delta x^6 }{ g_s l_s} \sim \frac{1}{e^2}\,,$$
respectively.

The field theory described above is valid in the limit in which we can honestly treat the vortices as stretched D$2$ branes. 
In other words, we
must be able to  neglect the effect of the junctions between the D$2$ and D$4$ branes.
A D$2$ brane terminating on  D$4$ can   be described as a spike of D$4$. The profile of the spike is $\propto l_s^2 / r$. 
The  decoupling  of  the junction happens for sufficiently large $
\Delta x^9 \gg l_s$,  so that the junction is very small, namely,
\beq
\frac{1}{g_s} \ll \delta_9 \,.
\label{ptpbad}
\eeq
This  assures, in particular,   the gauge coupling $g \sim (\sqrt{\delta_9} l_s)^{-1}$  to be smaller than the string scale.
We see a conflict between the two validity limits, (\ref{ptpbad}) on the one hand and (\ref{wachie}) -- (\ref{tbadd}),
on the other. Indeed, (\ref{tbad}) (with
$ g_s  \to  0$) is   the requirement that the scale of Higgsing in the bulk theory is smaller than the string scale.
It is obviously  incompatible with the constraint (\ref{ptpbad}). 

 Thus, the Hanany--Tong model
 on the world sheet  cannot
be obtained in the field-theoretic set-up.

\subsection{Some preliminary comments}
\label{dom4}

Here we pause to mention an issue which elucidates the distinctions 
between the two formulations: the D-brane and the soliton.

If the bulk theory is a  weakly coupled field theory (e.g. the model described in Sect.~\ref{wthse}, see \cite{SYrev}), the string thickness  is  $\ell \sim 1/(e \sqrt{\xi})$ with $e\ll 1$. It is parametrically larger than $1/\sqrt{\xi}$, the 
length scale set by the tension $T$ ($T\sim \xi$) because $e\ll 1$.
Under these circumstances, a weakly coupled sigma-model description for the translational modes is possible. 
The metric starts varying when the strings start to overlap in the perpendicular
plane. Thus, it is very smooth in  the tension scale. In other words, if we change the distance between the strings
by $\delta L \sim 1/\sqrt{\xi}$, the variation of the metric is negligible.

The D-brane description is, instead, completely different. 
The D-branes are infinitely thin objects, and
the low-energy physics is described by the massless 
open-string modes: a non-Abelian gauge theory with the translational modes in the adjoint representation.
The non-Abelian gauge symmetry is
spontaneously broken by the inter-brane distances. At large distances, the number of translational 
modes is $N$, as is the number of  branes. 
When the separation is zero, the number of massless models becomes $N^2$. This 
is a crucial difference with the sigma-model 
description of solitons where the dimension of the moduli space is never enhanced.

Let us consider two $\mathbb{CP}^{1}$ non-Abelian strings of thickness $\ell$, tension $T$ and relative distance $L$. 
Focus on the state in which these strings have the opposite $\vec{B}^3$ orientations (i.e. (1,0) + (0,1)).
In field theory we have in general $\ell \gg 1/\sqrt{T}$.
If we descend  to $L<\ell$, it is not possible to  describe separately the orientational moduli for the two strings. If the elementary strings, comprising the 2-string,
overlap in the transverse plane, the non-Abelian magnetic fluxes are summed up and
 the $\U(1)_3$ magnetic fluxes in the (1,1) string
should annihilate each other, with no relative orientation moduli   surviving.

 For a field-theoretic realization of the D-brane physics, we would need 
 $\ell \ll T^{-1/2}$. Then we could have,  simultaneously,
$\ell\ll L$ and    $LT^{1/2}\lsim 1$. If  $LT^{1/2}\lsim 1$
elementary strings in the 2-string configuration can be viewed as coinciding. At the same time,
the magnetic fluxes of the constituent strings do not overlap, because $\ell\ll L$.
Then, the configuration (\ref{Bfour}) would indeed be characterized by a well defined
set of independent orientational moduli. 
That is what we see in the D-brane description. 
This regime does not seem to be achievable  in weakly coupled
bulk theories.

The strategy we use in this paper is to take the HT model
{\em per se}, and then use it in the field-theory domain of validity.
 That is, 
 we consider the  sigma model obtained 
 upon integrating out the gauge fields of the HT model. This is the limit in 
 which the gauge fields becomes just   auxiliary fields.
Needless to say, 
this is not going to reproduce the ``exact'' sigma model that one could derive
in field theory, nor even  describe the D2-brane dynamics in the limit of validity (\ref{ptpbad}).
 But many features are hopefully captured. 
(For example, in the HT model the elementary
strings start interacting
when $L=1/(\sqrt{\xi} e_3)$, which is consistent with the bulk expectations,
see Eq. (\ref{omegamax})). The BPS sector lives up to this promise in full.

\subsection{Hanany--Tong model}
\label{whtmo}

As was mentioned, the
 bosonic sector of the HT model is  described by a U$(k)$
gauge field with field strength $F_{01}$; a complex scalar $\sigma$
in the adjoint of U$(k)$  (which correspond to the position of the D$2$ brane in the $x_{4,5}$ plane)
 in the same hypermultiplet as the gauge field;
a complex scalar $Z$ in the adjoint representation of U$(k)$
 (which correspond to the position of the D$2$ brane in the $x_{1,2}$ plane);
$N$ scalars $n_j$ in the fundamental of U$(k)$,
which we can combine in a $k \times N$ matrix $n_j^l$ (where
$j$ is a global SU$(N)$ index and $l$ is a gauge U$(k)$ index).

The parameters of the model are: (i) the two-dimensional U$(k)$ gauge coupling 
$g$ (with the dimension of a mass); (ii)
 the twisted mass $m_j$; (iii) the dimensionless  Fayet--Iliopoulos parameter $r$;
  and (iv) the theta angle $\theta$.
 (In the notation of Ref.~\cite{SYrev}, one has $r=2\beta$.)
 The FI parameter $r$ is not to be confused with $r=\sqrt{x_1^2+x_2^2}$
 which will not appear below.
 
 The classical value of the FI term $r$ is directly related to the four-dimensional gauge coupling,
\beq 
r=\frac{4 \pi}{e^2_{3}} \, .
\eeq
For each of the $N$
chiral multiplets $n_j$ one can introduce
a different twisted mass parameter $m_j$. Only the differences between 
the twisted masses are physically significant;
$\sum_{i=1,...,N} m_i$ can be set to zero by a linear shift in the
trace of $\sigma$.
Due to the chiral anomaly one can always set  the vacuum angle $\theta=0$
by virtue of a   phase rotation of the complex mass parameters $m_i$. 

The action of $\N=(2,2)$ U$(k)$ two-dimensional gauge model
 can be obtained by dimensional reduction of the four-dimension
$\mathcal{N}=1$ theory. The
standard conventions are summarized in \cite{witten}.  
The bosonic part of the Lagrangian takes the form
\beqn
&&
\frac{1}{g^2} \Tr \left( -\frac{1}{2} F^{\mu \nu } F_{\mu \nu} 
+\frac{1}{2} | {\mathcal D}_\mu \sigma |^2 - \frac{1}{8} ([\sigma,\sigma^\dagger])^2+\frac{1}{2} D^2
-g^2 \,  r \, D \right) 
\nonumber\\[3mm]
&& +\left( ({\mathcal D}^\mu n_i^\dagger) ({\mathcal D}_\mu n_i)-
\frac{1}{2} n_i^\dagger \{ \sigma-  \mathbb{I} \, m_i, \sigma^\dagger - \mathbb{I} \,  m_i^*  \} n_i
 + n_i^\dagger D n_i 
 \right)
 \nonumber\\[3mm]
&& + \Tr \left( | {\mathcal D}_\mu Z |^2 
-\frac{1}{2} \{\sigma,\sigma^\dagger \} \{ Z, Z^\dagger \} + 
(Z^\dagger \sigma Z \sigma^\dagger+  Z^\dagger \sigma^\dagger Z \sigma )
+Z^\dagger [D,Z]  \right) .
\nonumber\\
\label{ans44}
\eeqn
The symbol $\mathbb{I}$ is used for the $k \times k$ identity matrix.
The scalar fields in this action have the following dimensions:
\[ 
Z, \, n_i \propto [{\rm mass}]^0 \, , \qquad
D \propto [{\rm mass}]^2 \, , \qquad \sigma \propto [{\rm mass}] \, .
\]
The eigenvalues of $Z$ correspond to the 
 positions of the component strings in the perpendicular plane,
measured in the units of $1/\sqrt{T}$ where $T$ is the vortex tension.
The trace of $Z$ is completely decoupled from dynamics; therefore, we can (and will)
 set it to zero.\footnote{In terms of the parameter $L$ used previously,
 $2|z| = L\sqrt{T}$, see Eq.~(\ref{tfiga}).}

The classical vacua
are given by the condition of vanishing of the $D$-terms,
\beq 
D=-g^2 \left( [Z,Z^\dagger]+n n^\dagger - \mathbb{I} \, r \right) = 0 \, .
\label{tvand}
\eeq
For $m_i=0$ this constraint gives us the classical  moduli space. 
If the adjoint field $Z$ were not present,
the theory would  correspond to the gauged formulation of the $\N=(2,2)$ sigma
model with target space in the Grassmannian space 
\beq 
G_{N,k}=\frac{{\rm U}(N)}{{\rm U}(N-k) \times {\rm U}(k)} \, .
\label{tgrassm}
\eeq
The $Z$ field introduces new degrees of freedom in the Lagrangian
making the sigma model at hand more contrived.

The eigenvalues of $Z$ are the classical moduli   
 which must survive switching on quantum corrections.
In the limit when the difference  between the eigenvalues of $Z$ 
is $\gg 1$ 
the U$(k)$ gauge group is Higgsed to U(1)$^k$.
The adjoint field $Z$  is then decoupled, and
 we recover 
$k$ copies of  the supersymmetric sigma model with the target space
\beq
\mathbb{CP}^{N-1}=\frac{{\rm U}(N)}{{\rm U}(N-1) \times {\rm U}(1)} \, .
\label{ttargsp}
\eeq

In the opposite limit, in which 
 the eigenvalues of ${Z}$ fuse at a common value $z_0$,
the corresponding dynamics is  richer and more interesting.
In this limit  the matrix $Z$ can be put in a triangular form (with nonvanishing elements at the
main diagonal and above it).
Both diagonal entries are $z_0$. 
The degrees of freedom corresponding to the
upper-triangle elements of $Z$ are classically massless and 
couple nontrivially to  other degrees of freedom
of the  U$(k)$ theory.

At the quantum level the Fayet--Iliopolous term $r$, 
which determines the strength of interaction on the world sheet,
runs logarithmically at one
loop; by dimensional transmutation it is traded for a dynamical scale $\Lambda_{1+1}$
(see Sec.~\ref{rg}).
For $k=1$ this corresponds to the running  coupling 
of the asymptotically free $\mathbb{CP}^{N-1}$ sigma model.
In what follows we limit ourselves to $N=k=2$.
In order to study the system at weak coupling
we introduce the twisted  mass term 
\beq 
m_1=-m_2=m \, , \qquad | m |\gg \Lambda_{1+1}\,.
\label{ttm}
\eeq
For our purposes it is sufficient to assume $m$ real.

\subsection{ Moduli Space}
\label{tvms}

For $N=2$ and $k=2$, we can use the gauge fixing
\beq
 Z= \left(\begin{array}{cc}
z  & r^{1/2} \, \omega\, e^{i \zeta}  \\
0  & -z  \\
\end{array}\right) \, , \qquad n=\left(\begin{array}{cc}
a_1  &  a_2  \\
b_1  & b_2  \\
\end{array}\right) \, , 
\label{tfiga}
\eeq
where $\omega$ is a real positive parameter.
This does not completely fix  the gauge;
it remains to fix continuous U(1)'s,
\beq 
{\rm U}(1)_1 \, {\rm :} \quad U= \left(\begin{array}{cc}
e^{i \varphi}  &  0 \\
0  &   1 \\
\end{array} \right) \,, \qquad
{\rm U}(1)_2 \, {\rm :} \quad U=\left(\begin{array}{cc}
1  &  0 \\
0  &   e^{i \varphi} \\
\end{array}\right)  ,
\label{t27}
 \eeq
under which $z$ is uncharged, $$\tilde{\omega}=\omega e^{i \zeta}$$
 transforms as $(1,-1)$, $\, a_i$ as $(1,0)$
and $b_i$ as $(0,1)$. 
There is also some discrete subgroup of the gauge to fix.
With this parametrization, the $D$-term constraints have the form
\beq
 \sum_i |a_i|^2 = r \, (1-\omega^2) \, , \qquad
\sum_i |b_i|^2= \, r(1+\omega^2) \, , \qquad 
a_1 b_1^* +a_2 b_2^*=2 \sqrt{r} \, z^* \omega \, .
\label{tdtc}
\eeq
It follows that for fixed $|z|$ the allowed  range 
for $\omega$ is
\beq 
0 \leq \omega \leq
 \omega_{\rm max}=\sqrt{\frac{\sqrt{r^2+4 |z|^4}-2 |z|^2}{r}} \, . 
 \label{omegamax} 
 \eeq
The value of $\omega_{\rm max}$ gives us the measure of how
much the two elementary strings interact with each other.
 In the limit of $|z| \rightarrow \infty $,
$$
\omega_{\rm max} \approx \sqrt{r}  /|z| \,.
$$ 
In this limit $a_i$  and $b_i$ parametrize two
decoupled $\mathbb{CP}^1$'s with radii $\sqrt{r}$.
In order for the two copies of $\mathbb{CP}^1$
to interact, $z$ 
should be of the same order of magnitude as
$\sqrt{r}$. 
 This is completely consistent
 with what we expect from the bulk theory in the weakly coupled limit:
we know that the string thickness
 is of the order of
 \beq 
 \sqrt{\frac{r}{T}} \propto \sqrt{\frac{1}{\xi}} \frac{1}{e_3} \, . 
 \eeq
 
It is straightforward to check that the corrections to 
the metric of the two decoupled $\mathbb{CP}^1$'s
 for large $z$ are proportional to $1/z^2$;
this is inconsistent with what we expect from the four-dimensional 
gapped bulk theory in which these corrections should fall off
exponentially. 

The opposite limit $z=0$ corresponds to 
the requirement of orthogonality  of the
vectors $a_i$ and $b_i$. 
In this case 
$$
0 \leq \omega \leq 1\,.
$$
The section with $\omega^2=1$ corresponds to
a $\mathbb{CP}^1$ submanifold (the orientational moduli
of the component strings are aligned  in
the group space).
The section with $\omega^2=0$ corresponds to a point
(the component strings'   orientations in the group space are antiparallel).
At $z=0$ we use  the following gauge fixing:
\beqn
 a_i
 &=& r^{1/2}
 \sqrt{1-\omega^2} \, (\cos \alpha, e^{i \beta} \sin \alpha) \, ,
 \nonumber\\[3mm]
b_i
&=& r^{1/2}
\sqrt{1+\omega^2} \, ( e^{-i \beta} \sin \alpha, - \cos \alpha) \, .
\label{tfgf}
\eeqn
As a result, the matrix $Z$ takes a very simple  form
\beq 
Z= \left(\begin{array}{cc}
0  & \sqrt{r} \, \omega \, e^{i \zeta}  \\[1mm]
0  & 0  \\
\end{array}\right) \, .
\label{tvsf}
 \eeq
The orientational  moduli are encoded in the real parameter $\omega$
and  three angles, $(\zeta,\alpha,\beta)$.

\subsection{Kinetic term}
\label{tkintt}

In order to get the metric on the moduli space, we have to find
the saddle-point value of the
gauge field $A_{\mu}$ and plug it back in the Lagrangian. 
We  work in the limit of coincident strings, $z=0$.

With our gauge choice  a straightforward calculation gives
\beqn
 A_\mu^0
 &=&
 \frac{2 \omega^4 \left[(\partial_\mu \zeta)-2 \sin^2 \alpha (\partial_\mu \beta)\right]}
{1+2  \omega^2-\omega^4} \, , 
\nonumber\\[3mm]
A_\mu^3
&=&
\frac{2 \left[\sin^2 \alpha \, (1-\omega^4) (\partial_\mu \beta) +r \omega^2 (\partial_\mu \zeta)\right]}
{1+2  \omega^2-\omega^4} \, ,
\nonumber\\[3mm]
A_\mu^1
&=&
-2 \sqrt{\frac{1-\omega^2}{1+\omega^2}}
\left[ \sin \beta \, (\partial_\mu \alpha)+ \sin \alpha \, \cos \alpha  \, \cos \beta \, 
(\partial_\mu \beta)
\right] \, ,
\nonumber\\[3mm]
 A_\mu^2
 &=&
 2 \sqrt{\frac{1-\omega^2}{1+\omega^2}}
\left[ - \cos \beta \, (\partial_\mu \alpha)+ \sin \alpha \, \cos \alpha  \, \sin \beta \,
(\partial_\mu \beta )\right] \, , 
\label{tspv}
\eeqn
where 
\[ A_\mu= \frac{A_\mu^0 \, \mathbb{I} + A_\mu^k \, \sigma_k}{2} \, ,\]
and $\sigma_k$ are the Pauli matrices.
To find the moduli space metric we have
to substitute these expressions in the kinetic term,
\beqn
 &&
 r \left(  \, \frac{1+2 \omega^2 -\omega^4 }{1-\omega^4} (\partial_\mu \omega)^2 +
 2 \omega^2 \left( (\partial_\mu \alpha)^2 +\left( \frac{\sin 2 \alpha}{2}\  \partial_\mu \beta \right)^2 \right) + \right.
 \nonumber\\[3mm]
 &&\left.  + \frac{\omega^2(1-\omega^4)}{1+2  \omega^2-\omega^4}
(\partial_\mu \zeta- 2 (\sin^2 \alpha) \partial^\mu \beta)^2 \right)  \, .
\label{tsppv}
\eeqn
The term proportional to $(\partial_\mu \omega)^2$ diverges at $\omega=1$.
Luckily this is not a bad divergence. It can be eliminated by virtue of  a change of variables.
Indeed,
 define
\beq 
\kappa=\sqrt{1-\omega}\, , \qquad \omega=1-\kappa^2 \, .
\label{kappona} 
\eeq
Then the relevant piece of the metric is
\beq 
4 r \, A \,
(\partial_\mu \kappa)^2 \, ,\qquad 
 A=\frac{\kappa^8 -4  \kappa^6 + 4  \kappa^4-2  }
{\kappa^6 -4  \kappa^4 + 6  \kappa^2 -4 } \, .
\eeq
It is completely smooth at $\kappa=0$ (which corresponds to $\omega=1$
in the previous choice of variables).

\subsection{Some topology}
\label{tsoto}

The coordinates  in the moduli space that we have introduced
vary in the following intervals:
\beq 
0 \leq \omega \leq 1 \, , \qquad 
0 \leq \alpha \leq \frac{\pi}{2} \, , \qquad 
 0 \leq \zeta \leq 2 \pi \,, \qquad  
 0 \leq \beta \leq 2 \pi \, .
 \label{tinter}
\eeq
First we will consider sections at generic values of $\omega \neq 0,\,\, \sqrt{r}$.
We can  pass to an alternative gauge fixing, 

\beqn
 a_i
 &=& r^{1/2} \,
 \sqrt{1-\omega^2} \, (\cos \alpha, e^{i \beta} \sin \alpha) \, e^{-i \zeta/2}\, , 
  \nonumber\\[3mm]
b_i
&=&  r^{1/2} \,
\sqrt{1+\omega^2} \, ( e^{-i \beta} \sin \alpha, - \cos \alpha) \, e^{+i \zeta/2} \, ,
 \nonumber\\[3mm]
 \tilde{\omega}
 &=&
\omega\,.
 \label{taltgf}
 \eeqn
The point with the coordinates $(\omega,\alpha,\beta,\zeta)$
is then identified with the point with the coordinates  $(\omega,\alpha,\beta,\zeta+ 2 \pi)$.
The topology of the sections at constant $\omega$ is then 
given by $S^3 / \mathbb{Z}_2$. This is due to the fact that
 the point $(a_i,b_i)$ is identified with $-(a_i,b_i)$.
At $\omega=0$ the section is given by just a point.
At $\omega=\sqrt{r}$ the section is given by  $S^2=\mathbb{CP}^1$,
parametrized by $(\alpha,\beta)$.
The topology of the moduli space is $\mathbb{CP}^2/{Z}_2$.

\subsection{Twisted mass term}
\label{ttmtt}

To warm up we start with the simple  case of the elementary string,
$k=1$. Then 
we can choose the gauge in such a way that
\beq 
n_1=\cos \alpha \, , \qquad n_2=e^{i \beta} \sin \alpha \, ,
\label{tngc}
\eeq
where $(\alpha,\beta)$ parametrize the $\mathbb{CP}^1$ moduli.
To find the mass-term-generated effective potential  
we  integrate out $\sigma$. The only nonvanishing part of the potential is
\beq 
V=  \sum_i n_i^\dagger (\sigma-m_i) (\sigma^*-m_i^*) n_i 
\label{tnonvpp}
\eeq
implying the following  saddle-point value of $\sigma$:
\beq 
\sigma=m  (\cos^2 \alpha -\sin^2 \alpha) \, .
\label{tspvs}
\eeq
Substituting (\ref{tspvs}) in (\ref{tnonvpp}) we get
 \beq V= m^2  \, r \, \sin^2 (2 \alpha) \, .
 \label{ttpot}
 \eeq
This is the standard twisted mass term in the $\N=(2,2)$
$\mathbb{CP}^1$ sigma model.

\begin{figure}[h]
\begin{center}
\leavevmode
\epsfxsize 6 cm
\epsffile{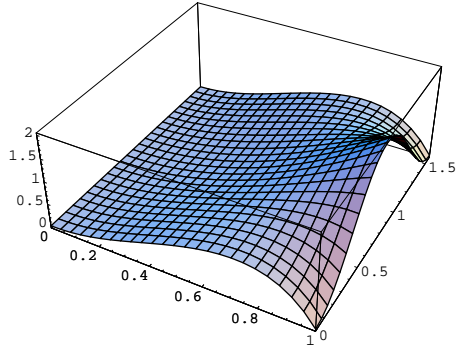}
\end{center}
\caption{\footnotesize Potential as a function of $\omega$ and $\alpha$.}
\label{led}
\end{figure}

After this successful exercise we turn to
 the $k=2$ case. For 2-strings we
 have to determine $\sigma$ from the   potential
\beqn 
V
&=&  \frac{1}{8} ([\sigma,\sigma^\dagger])^2 + 
\frac{1}{2} n_i^\dagger 
\{ \sigma-\mathbb{I} \,m_i, \sigma^\dagger - \mathbb{I} \, m_i^*\} n_i 
\nonumber\\[3mm]
&+&
\frac{1}{2} \{\sigma,\sigma^\dagger \} \{ Z, Z^\dagger \} - 
(Z^\dagger \sigma Z \sigma^\dagger+  Z^\dagger \sigma^\dagger Z \sigma ) \, .
\label{tanpot}
\eeqn
Integrating out $\sigma$, we arrive at
\beq 
\sigma=m \,  \left( \begin{array}{cc}
\frac{(1-3 \omega^4) \cos 2 \alpha }{1+2  \omega^2 -\omega^4}
  & \frac{e^{i \beta} \sqrt{1-\omega^2} \sin 2 \alpha }{\sqrt{1+\omega^2}}  \\[3mm]
\frac{e^{-i \beta} \sqrt{1-\omega^2} \sin 2 \alpha }{\sqrt{1+\omega^2}}  
 & -\frac{(1+ \omega^4) \cos 2 \alpha }{1+2  \omega^2 -\omega^4}  \\[2mm]
\end{array} \right) \, .
 \label{ssigma} 
 \eeq
With this saddle-point value of $\sigma$  the potential takes the form
\beq
 V=m^2  \, r \, \frac{\omega^2 (3  + 2  \omega^2 - 3 \omega^4
+(1-2  \omega^2-\omega^4) \cos 4 \alpha
) }{1+2  \omega^2-\omega^4} \, .
\label{tsadpop}
\eeq
It depends only on $\omega$ and $\alpha$. A plot of the potential 
(\ref{tsadpop}) is displayed in Fig.~\ref{led}.

Note that in this plot the line $\omega=0$ corresponds to a single point in the
moduli space (the $(1,1)$ string).
At $\omega^2=1$ the potential reduces to
$$V=2 m^2 \, r \, \sin^2 2 \alpha\,,$$
 exactly twice the potential on the elementary string (cf. Eq.~(\ref{ttpot})).

\section{Spectrum of excitations}
\label{spoex}

\subsection{Perturbative excitations}
\label{ptpeex}

After the potential on the 2-string world sheet is found,
we can  compute the mass of the perturbative excitations near
each of three vacua. 
Let us start from the $(1,1)$ string, which corresponds to $\omega=0$
and $\kappa=1$ (the minimum on the left-hand side in Fig.~\ref{led}).
 The  mass-squared of the excitations
is given by
\beq 
M^2=\frac{\partial^2_{\kappa,\kappa} V}{4 A r}=2 m^2 (3+\cos 4 \alpha) \, .
\label{tmse}
\eeq
There are  two normal modes, one at $\alpha=0$ and another at $\alpha=\pi/4$.
Thus, there are two scalar excitations with mass  
$2\sqrt{2} m$ plus two scalar excitations with mass   $2 m$
(and their superpartners, of course).

For the $(2,0)$ string (at $\omega=1$ and  $\kappa=0$)
 the situation is slightly different. 
The oscillations can be both in the $\alpha$ and   $\kappa$ coordinates.
The mixed term  $\partial^2_{\kappa,\alpha} V$ vanishes.
The mass of each of these excitations is
\beq M_\kappa^2=\frac{\partial^2_{\kappa,\kappa} V}{4 A r}=
8 m^2  \, , \qquad 
M_\alpha^2=\frac{\partial^2_{\alpha,\alpha} V}{2 r \omega^2}=
8 m^2 \, .\eeq 
So there are a total of four scalar states with masses $2\sqrt{2} m$
(plus their superpartners).  It is, of course,  the same for the $(0,2)$ string.

\subsection{The BPS-saturated kinks}
\label{kkki}

\begin{figure}[h]
\begin{center}
\leavevmode
\epsfxsize 6 cm
\epsffile{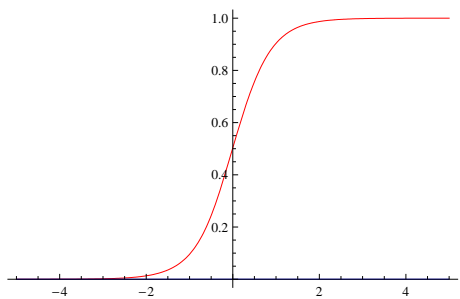}
\end{center}
\caption{\footnotesize The profile function $\kappa(x)$ for the elementary 
kink between the $(2,0)$ and   $(1,1)$ 2-strings.
}
\label{kink1}
\end{figure}

For the elementary kink
(which interpolates between the vacuum at $\omega=\sqrt{r}$, $\alpha=0$ 
and the vacuum at $\omega=0$ and has the unit magnetic flux),
 we can choose the ansatz 
$\alpha=\beta=\zeta=0$ and introduce a profile function $\kappa(x)$.
Using the variable (\ref{kappona}),
the energy functional for this kink can be written as
\beq 
{\mathcal E}=
\int d x \, r \, \left[ 4   \,A \, 
(\partial_x \kappa)^2 + \frac{4 m^2 \kappa^2}{A}  (1-\kappa^2 )^2
 \right]   \, ,
 \label{tenfuek}
 \eeq
 where $x$ is the coordinate along the 2-string axis, and the boundary conditions
 on $\kappa$ are
 \beq
\kappa (x= - \infty ) =0  \,,\qquad \kappa (x=+\infty ) = 1 \,.
 \label{tboco}
 \eeq
The Bogomol'nyi completion is straightforward,
\beq 
{\mathcal E}=
\int d x \, r \left[
\left( 2  \sqrt{A} (\partial_x \kappa ) \pm
\frac{(2 m \kappa)(1-\kappa^2)}{\sqrt{A}}
\right)^2 \mp 2 m r \, \partial_x \left( 2   \kappa^2-\kappa^4\right)
\right].
\label{tbogc}
\eeq
For BPS (elementary) kinks
one must have
\beq
 2  \, \sqrt{A} (\partial_x \kappa ) \pm
\frac{(2 m \kappa)(1-\kappa^2)}{\sqrt{A}}
 =0\,.
\label{bpsele}
\eeq
If this equation is satisfied (and it is, see below) the tension of the elementary kink is 
\beq 
T_{(2,0)\to (1,1)}=2 m r \, . 
\label{tmek}
\eeq
The solution to Eq.~(\ref{bpsele}) with the boundary conditions (\ref{tboco}) can be 
found numerically 
(see Fig.~\ref{kink1}).

Now, we can  consider a composite kink, interpolating
between the $(2,0)$ and the $(0,2)$ strings.
In our notation this corresponds to an interpolation
between the vacuum at
 $\alpha=0$, $\omega=1$
and the one at $\alpha=\pi/2$, $\omega=1$.
As we will show shortly, the mass of the composite BPS-saturated kink is 
$4 m r$, twice larger than in Eq.~(\ref{tmek}). This means that
there is no interaction between the elementary kinks $(2,0)\to (1,1)$ and $(1,1)\to (0,2)$
comprising the $(2,0)\to (0,2)$ kink.
Hence, the relative distance between the component  elementary kinks is a modulus.
 
The simplest solution (one of a family) can be found keeping $\omega$ constant.
The energy functional then reduces to that  given by the sine-Gordon model,
\beq
{\mathcal E}= 
\int  d x \, \left[ 2 r (\partial_x \alpha)^2 + 2 m^2 r \sin^2 (2 \alpha)
\right] \, .
\eeq
The Bogomol'nyi completion is
\beq 
{\mathcal E}=\int  d x \, \left\{ \left( \sqrt{2 r} (\partial_x \alpha) \pm \sqrt{2 r} \, m \, \sin (2 \alpha) \right)^2
\pm \partial_x \left(  2 r m \cos (2 \alpha)
\right)
\right\} \, .
\eeq
Assuming that
\beqn
&&
\sqrt{2 r} (\partial_x \alpha) \pm \sqrt{2 r} \, m\,  \sin (2 \alpha) =0\,,
\nonumber\\[2mm]
&&
\alpha (x= - \infty ) =0  \,,\qquad \alpha(x=+\infty ) = \frac{\pi}{2} \,,
\label{tabc}
\eeqn
we find the 
tension  
\beq 
T_{(2,0)\to (0,2)}=4 m r  = 2T_{(2,0)\to (1,1)} \, . 
\label{22tens}
\eeq
Next, we have to check that the first-order equation ~(\ref{tabc}) does have solutions.
\begin{figure}[h]
\begin{center}
\leavevmode
\epsfxsize 6 cm
\epsffile{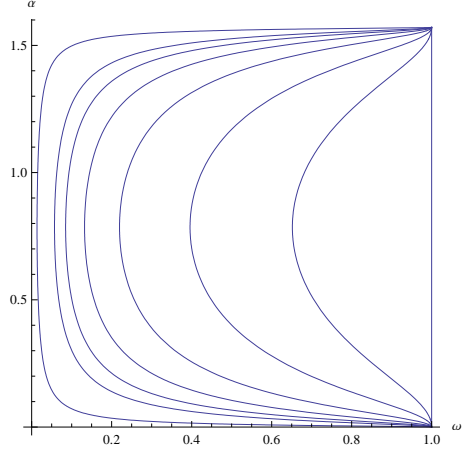}
\end{center}
\caption{\footnotesize The family of degenerate composite kinks 
(interpolating between the $(2,0)$ and the $(0,2)$ strings)
in the $(\omega,\alpha)$ plane.
The line at $\omega=1$  corresponds to the kink with the smallest thickness.
In the large thickness limit the solution degenerates in two elementary  kinks at an (almost)
infinite distance. 
}
\label{kinks}
\end{figure}
To find the most general solution  we have to introduce two
 profile functions now, $\alpha(x)$ and $\kappa(x)$, determining the energy functional
\beq 
{\mathcal E}=
\int d x \, \left[ 4 r   \,A \, 
(\partial_x \kappa)^2 + 2 r (1-\kappa^2)^2 (\partial_x \alpha)^2+ V
 \right] \, ,
 \label{tdetef}
 \eeq
where
\beq V= 2 \, r \, m^2 \, (\sin^2 2 \alpha )\, (1-\kappa^2)^2 +
\frac{4 m^2 \, r  \, (\cos^2 2 \alpha ) \, \kappa^2 \, (1-\kappa^2)^2}{A} \, .
 \eeq
The generic Bogomol'nyi completion takes the form
\beqn
&&{\mathcal E}
=
\int d x \, r \, 
 \left\{
\left( 2 \sqrt{A} (\partial_x \kappa ) \pm
\frac{(2 m \kappa)(1-\kappa^2)(\cos 2 \alpha)}{\sqrt{A}}
\right)^2 + \right.  
\nonumber\\[3mm]
&& \left. +\left( \sqrt{2} (1-\kappa^2) (\partial_x \alpha \mp m \sin 2 \alpha)
\right)^2  \mp 2 m \partial_{x} \left(  (1-\kappa^2)^2  \cos 2 \alpha \right) 
 \right\} \, .
\nonumber \\
 \label{genericB}
 \eeqn
 The BPS equations are
 \beqn
 &&
 2 \sqrt{A} (\partial_x \kappa ) \pm
\frac{(2 m \kappa)(1-\kappa^2)(\cos 2 \alpha)}{\sqrt{A}}
=0\,, 
\nonumber\\[3mm]
&&  \partial_x \alpha \mp m \sin 2 \alpha
=0\,.
 \eeqn
 They can be solved numerically, as shown in Fig.~\ref{kinks} in the $(\omega,\alpha)$ plane.
 Needless to say, the mass of every solution in this family obeys Eq.~(\ref{22tens}).

\subsection{R symmetries}

The $\N=(2,2)$ U$(k)$ theory has some interesting $R$-symmetries, which are the same as in 
the $k=1$ case \cite{witten,sy2009}.
 Let us denote  the superpartners of
 $n_i$ and $Z$ by  
$(\psi^{n_i}, \psi^{Z})$;  $\lambda$ is the world-sheet gaugino.
 There exists a vectorial symmetry which  acts only on the following fermions:
 \beq  \psi^{n_i}_{L,R} \rightarrow  e^{i \gamma} \psi^{n_i}_{L,R} \,  , \qquad
  \psi^{Z}_{L,R} \rightarrow  e^{i \gamma} \psi^{Z}_{L,R} \, , \qquad
  \lambda_{R,L} \rightarrow e^{-i \gamma} \lambda_{L,R} \, . 
 \eeq
This classical symmetry is  unbroken by quantum effects
and  unbroken by the twisted mass term.

In addition, in the limit of vanishing twisted masses,
there is   an axial U(1) symmetry
which is broken to $Z_{2 N}$ by the quantum (chiral) anomaly, 
\beqn
 &&
 \psi^{n_i}_{L} \rightarrow  e^{- i \gamma} \psi^{n_i}_{L}  \, , \qquad
 \psi^{n_i}_{R} \rightarrow  e^{ i \gamma} \psi^{n_i}_{R}  \, ,
\nonumber\\[3mm]
&&
 \psi^{Z}_{L} \rightarrow  e^{- i \gamma} \psi^{Z}_{L}  \, , \qquad
 \psi^{Z}_{R} \rightarrow  e^{ i \gamma} \psi^{Z}_{R}  \, ,
\nonumber\\[3mm]
&&
 \lambda_{L} \rightarrow  e^{- i \gamma} \lambda_{L}  \, , \qquad
 \lambda_{R} \rightarrow  e^{ i \gamma} \lambda_{R}  \, , \qquad
 \sigma \rightarrow e^{2 i \gamma} \sigma \, .
 \label{qchan}
 \eeqn
The twisted mass terms generically  break this symmetry.
However, with the particular choice
\beq 
m_i = m \left(e^{2 \pi i/N}, e^{4 \pi i/N }, \ldots , e^{2(N-1) \pi i/N}, 1 \right) \,
\label{znsymmc}
 \eeq
a discrete $Z_{2N}$ subgroup survives the inclusion of both
 the anomaly and  mass terms,
\beqn 
 &&
 \psi^{n_i}_{L} \rightarrow  e^{- i \gamma_k} \psi^{n_{i-k}}_{L}  \, , \qquad
 \psi^{n_i}_{R} \rightarrow  e^{ i \gamma_k} \psi^{n_{i-k}}_{R}  \, ,
\nonumber\\[3mm]
&&
 \psi^{Z}_{L} \rightarrow  e^{- i \gamma_k} \psi^{Z}_{L}  \, , \qquad
 \psi^{Z}_{R} \rightarrow  e^{ i \gamma_k} \psi^{Z}_{R}  \, ,
\nonumber\\[3mm]
 &&
 \lambda_{L} \rightarrow  e^{- i \gamma_k} \lambda_{L}  \, , \qquad
 \lambda_{R} \rightarrow  e^{ i \gamma_k} \lambda_{R}  \, , \qquad
 \sigma \rightarrow e^{2 i \gamma_k} \sigma \, , 
\nonumber\\[3mm]
&& n_i \rightarrow n_{i-k}  \, , \qquad \gamma_k=\frac{\pi k}{2N} \,  \, \, \, {\rm with} \, \, \, \, k=1, \ldots ,2N \, .
\label{survi}
\eeqn
In the special case $k=N=2$ under consideration,
we   choose $m_1=-m_2=m$. As a result, there is a discrete $\mathbb{Z}_4$
symmetry. 

From Eq. (\ref{ssigma}) we can check that for the $(2,0)$ vacuum
$\sigma_0 \neq 0$ and $\vec{\sigma}=0$ while
 for the $(1,1)$ vacuum $\sigma_0=0$
and $\vec{\sigma}\neq 0$. A VEV for $\sigma_0$  spontaneously breaks 
$\mathbb{Z}_4$ to $\mathbb{Z}_2$, while a VEV for $\vec{\sigma}$ does 
{\em not} break
the $\mathbb{Z}_4$ symmetry at all, because the phase can be eliminated by a gauge transformation.
Hence, the discrete $\mathbb{Z}_4$  symmetry is spontaneously broken to $\mathbb{Z}_2$
in the $(0,2)$ vacuum. It is unbroken in the $(1,1)$ vacuum.

\subsection{A general perspective}
\label{tsum}

The sigma model 
on the 2-string world sheet is quite unconventional; the moduli space
 is not a homogeneous space
and its topology, that of $\mathbb{CP}^2/{Z}_2$, is rather weird.
At $m=0$  the physics  described by this model  is strongly coupled
and hard to work with.

On the other hand, in the limit $m\gg\Lambda_{1+1}$ 
we are at weak coupling and can study the problem
in a (quasi)classical way. We found three vacua
which we can be identified with the $(2,0)$, $(0,2)$ and $(1,1)$
strings of the four-dimensional theory.

In the $\N=(2,2)$ theory, because of the Witten index,
the number of vacua should not change as a function
of $m$. Therefore, we conclude that the theory has three
vacua not only at large $m$, but also in the $m\rightarrow 0$ limit.

We see that two of these three vacua
(which correspond to the
$(2,0)$ and the $(0,2)$ strings in the $m\gg \Lambda_{1+1}$ limit )
 spontaneously break 
the anomaly-free $\mathbb{Z}_4$ symmetry 
of the model down to $\mathbb{Z}_2$.
The third vacuum (which  corresponds to the
$(1,1)$ vortex in the $m \gg \Lambda_{1+1}$ limit)  leaves this symmetry unbroken.
This implies, in turn, that in the
latter vacuum   the fermionic 
condensate must vanish. This is an important finding.

The BPS kinks interpolating
between   various pairs of vacua which we found correspond to monopoles of
the four-dimensional theory.
It is remarkable that in all three cases the masses of the kinks are exactly equal to the
masses of the 't Hooft--Polyakov monopole (and double monopole in the third case)
on the Coulomb branch of the bulk theory. This is exactly the phenomenon first observed in
\cite{SYmon}. It lends credence to the HT model as the theory correctly describing
the BPS sector in the composite strings.

It should be possible to
study dyonic kink. Moreover, for $k$-strings with $k>2$
we should be able to see kinks describing confined monopoles with the magnetic charges
1,2, ..., up to $k$.

\section{Composite strings at $m\to 0$}
\label{costr}

\subsection{Quantum moduli space}
\label{quantum}

The problem 
of complete characterization of the quantum moduli space for 2-strings
 is quite complicated;  no final solution is known at the moment. 
 However, our previous analysis of the $m
\neq 0$ case
provides us with some hints which we would like to summarize here.
If $m\to 0$ the potential vanishes, and we are left with the sigma model dynamics.

When we speak of the elementary  non-Abelian strings,  the translational sector is decoupled, 
and we can consider the $\N=(2,2)$  ${\mathbb{CP}^{N-1}}$  sigma model living on 
the world sheet of
an infinite straight string. 
In composite strings, even if we restrict ourselves
to the low-energy approximation, we cannot decouple the translational sector from the orientational one. 
Only the overall translational coordinate can be factored out, while the relative translations
 are inevitably entangled with the orientational modes.

Thus, we have to quantize a theory  of entangled  moduli,    some of them are noncompact   (the relative positions) 
while others are compact  (the   orientational moduli).
 The classical moduli space of $k$ non-Abelian elementary
strings in the bulk theory with $N$ colors and $N$ flavors will be referred to as   ${\cal M}_{k,N}$.
The real dimension of this moduli space is $2 k N$. For well separated constituent strings, this moduli space decomposes into the product of $k$ distinct factors ${\mathcal M}_{1,N} = {\mathbb{CP}^{N-1}} \times C$, modulo  permutation 
group $S_k$. 

Intuition obtained in  the elementary-string problem teaches us that 
quantum effects have a very different impact on the compact and noncompact parts of the moduli space. Sigma models
on the 
compact manifolds, generically, are subject to strong-coupling effects 
and develop a mass gap --  only a discrete number of vacuum states survives. 
Noncompact directions, instead, survive in the infrared as genuine moduli. Thus, we expect that
in the 2-string
 problem the quantum vacuum manifold will be spanned on the moduli
 describing  relative position of the elementary strings and will consist of a few sectors
labeled by appropriate  fermion condensates. That is the quantum counterpart of Fig.~\ref{4vs3}.
Since the problem is defined in $1+1$, there are also long-range logarithmic fluctuations of the non-compact moduli to be considered (see Sect.~\ref{tf}).

One can apply the following strategy:
 fix the spatial distance between the constituent  strings
 and then quantize the compact manifold   obtained in this way. 
Then vary the distance adiabatically. Finally, check 
whether or not
quantum fluctuation of the translational moduli (the non-compact part of the moduli space) 
alter the result.\footnote{We are grateful to  D.~Tong for pointing out to us the necessity of such a verification.}
The number of states we start from may be larger than the number of discrete moduli 
subspaces in which they are grouped. This was the case with $m\neq 0$.
Let us see how the vacua evolve as the distance varies from infinity to zero, 
in the specific example of ${\cal M}_{2,2}$. 

When the distance is large  $L \gg \ell$, we have to quantize two separate $ {\mathbb{CP}^{1}}$ models 
on the world sheets of  two separate strings. 
More exactly,  the
overall  theory is a   sigma model with the
target space $(\C \times {\mathbb{CP}^{1}} \times {\mathbb{CP}^{1}}) /Z_2$ where 
the $\Z_2$ factor is the exchange between two ${\mathbb{CP}^{1}}$'s and parity in  $\C$
(the relative position coordinate). 
This $\Z_2$ factor is crucial in what follows. 
At infinite separation we can quantize the two ${\mathbb{CP}^{1}}$'s separately, and 
then introduce the $\Z_2$ factorization,  at the level of the spectrum. 
Each string has two ground states where the wave function is spread uniformly around the ${\mathbb{CP}^{1}}$ 
manifold, while the (bi)fermion condensates are 
$\langle \bar\psi\psi \rangle = \pm \Lambda$. We call these ground states $|\pm\rangle_{1}$ and $|\pm\rangle_{2}$ respectively for the first and second strings. In total we have four states,
\beq
|+\rangle_1|+\rangle_2 \ , \qquad |+\rangle_1|-\rangle_2 \ , \qquad |-\rangle_1|+\rangle_2  \ , \qquad |-\rangle_1 |-\rangle_2 \ .
\label{stati}
\eeq 
Now we have to take into account the $\Z_2$ factor.
The first and fourth states are  invariant under the exchange $1 \leftrightarrow 2$. They, thus, 
belong to two separate manifolds ${\cal M}_{++}$ and ${\cal M}_{--}$. Since the exchange acts also on the relative position, the two manifolds are cones over the $ \S^1 / \Z_2$ angular variable.
The second and third states interchange under
$\Z_2$. That means that they belong to the same manifold ${\cal M}_{+-}$, which is asymptotically a cone over $\S^1$.
The two states are antipodal with respect to the angular variable.

The ground states of the 2-string are thus grouped exactly as in Fig.~\ref{4vs3}. There is a conceptual difference,
though. In the mass-deformed theory, the three manifolds are distinguished by the total, conserved, non-Abelian magnetic flux. In the $m=0$ case the the wave function is always spread uniformly around the ${\mathbb{CP}^{1}}$ manifolds, and thus the average non-Abelian magnetic flux vanishes for all of them. 
What  distinguishes them is the action of the residual
$\Z_4$  $R$-symmetry. This symmetry exchanges ${\cal M}_{++}$ and ${\cal M}_{--}$. 
Inside the manifold ${\cal M}_{+-}$ it acts as parity. Clearly, the central element of ${\cal M}_{+-}$ is the only state invariant under the residual $R$ symmetry.

As the distance 
between the elementary strings
becomes small enough,
one  can no longer quantize two ${\mathbb{CP}^{1}}$'s separately, in isolation from the  
translational part of the moduli space.  One can argue, however, that the number of 
the ground states must remain the same. In particular, at zero separation there are only three ground states.
The second and third in (\ref{stati}) must coalesce into a unique state, the central element of ${\cal M}_{+-}$.
The fourth vacuum state is not seen at zero separation.

\begin{figure}[h!t]
\epsfxsize=8cm
\centerline{\epsfbox{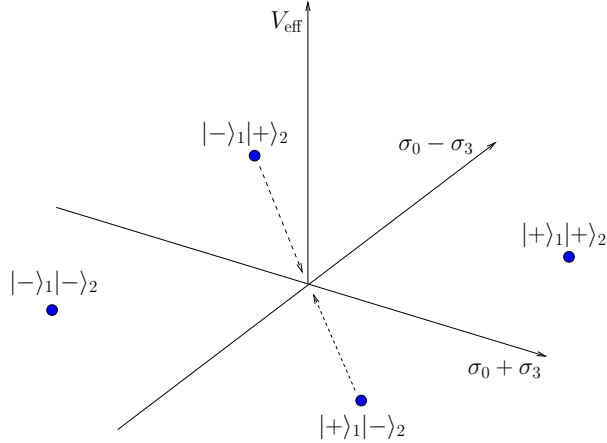}}
\caption{{\footnotesize The four vacua in the space of fermion condensate $\sigma$. As the separation goes to zero, two of them coalesce into a unique $\Z_4$-invariant state.}}
\label{three-vacua}
\end{figure}

In the HT model the separated strings imply an expectation value of
 $Z$ of the form
\beq
Z=d t_3 = \left(  
\begin{array}{ll}
L/2 & × \\[2mm]
× & -L/2                 \end{array} 
 \right) \ ,
 \label{dom8}
\eeq
which leads, in turn, to the  gauge group 
breaking,  $\U(2) \to \U(1) \times \U(1)$. 
In this language, the fermion condensate  is represented by
 the adjoint scalar field $\sigma$,
\beq
 \sigma = \bar\psi^{n_i} \psi^{n_i} =\sigma_0 \,\mathbb{I} +\vec{\sigma}\cdot \vec{\tau}\,,
 \label{dom8p}
\eeq
which is a member of the auxiliary gauge multiplet. 
If we could compute the quantum-generated effective potential 
$V_{\rm eff}(\sigma)$ for this scalar field, the problem is solved. 
At large separations $V_{\rm eff}(\sigma)$ reduces to
\beq
V_{\rm eff}(\sigma) = V_{\mathbb{CP}^{1}}(\sigma_0 + \sigma_3) + 
V_{\mathbb{CP}^{1}}(\sigma_0 - \sigma_3)\,,
\label{dom8pp}
\eeq
i.e.  the sum of   two ${\mathbb{CP}^{1}}$ effective potentials, which are, of course, known in the
literature.
The four vacua (\ref{stati}) can   be pictured in the space of fermion condensates, see Fig.~\ref{three-vacua}.

Now let us vary the separation and
try to  infer what happens with the  vacua at $L\to 0$. 
At $L=0$ the $\SU(2)$ symmetry is restored; hence, the effective potential must 
depend only on $\sigma_0$ and  $|\sigma|$.
Choosing the unitary gauge one can always set  $\vec\sigma$ in the third direction. 
Conservation of the number of vacua, together with the symmetry $\sigma_0 \to - \sigma_0$,
implies that the second and third vacua in (\ref{stati}) must coalesce  at $\sigma_0 =0$.
Invariance under $Z_4$ implies $\sigma_3 =0 $.
This state is topologically equivalent to the ANO string. See Sect.~\ref{tf} for a
 discussion of transversal fluctuations.

Summarizing, the quantum effective potential for $\sigma$, at zero separation, must
have  three vacua at $\sigma_0 =\pm \Lambda$ and $\sigma_0 =0$. 
These are the quantum analogs of three states  $(0,2)$, $(2,0)$, and $(1,1)$ in the mass-deformed theory.

\subsection{Instantons}
\label{instantons}

Now we will address another topological aspect of the HT model, namely instantons.
Their role is important. By virtue of the index theorem
they generate fermion zero modes, which, in turn, in conjunction with $\N=(2,2)$ supersymmetry
lead to bifermion condensates
(for a review see e.g. \cite{nsvz}).

We again fix the separation $L$, and consider quantization of the compact manifold   obtained at given $L$. 
The topology of this manifold, for $L \neq 0$, is the same 
as that of  ${\mathbb{CP}^{1}} \times {\mathbb{CP}^{1}}$. 
The existence/nonexistence of instantons is determined by the second homotopy group of the manifold,
$\pi_2({\mathbb{CP}^{1}} \times {\mathbb{CP}^{1}}) =\Z \oplus \Z$, and, thus, we have two distinct 
winding numbers, one for each ${\mathbb{CP}^{1}} $. At $L=0$ the topology is that of
$\mathbb{CP}^{2}/{Z}_{2}$. 
Defining $\mathbb{CP}^{2}$ as the identification 
$$
(z_1,z_2,z_3) \simeq (\lambda z_1, \lambda z_2, \lambda z_3)\,,
$$ 
the $\Z_2$ action is $(z_1,z_2,z_3) \to (z_1,-z_2,-z_3)$. 
The ANO string corresponds here to the fixed point of the orbifold $(1,0,0)$.
Other fixed points are the $\mathbb{CP}^{1}$ submanifold defined by $z_1=0$.
Note that the metric does not coincide exactly with that of $\mathbb{CP}^{2}/{Z}_{2}$.
However, for the purpose of discussion of the  instanton numbers and their zero modes, the result is the same.

The drastic change of topology 
in passing from $L\neq 0$ to $L=0$
 affects the instanton number which becomes  $\pi_2 (\mathbb{CP}^{2}/{Z}_{2}) = \Z$ where $\Z$ is 
is in one-to-one correspondence with the relative orientation. For example, the $(1,0)$  and  
$(0,-1)$ instantons, at $L=0$ merge into a unique topological sector.\footnote{ 
The notation used above to mark instantons is self-evident.}
They are two elements of the instanton  moduli space, obtained by the action of the $\SU(2)$ symmetry between the coordinates $z_2$ and $z_3$ of the orbifold. The cycle $(1,1)$ becomes   contractible at $L=0$.

The instanton moduli space  for $\mathbb{CP}^{1}$ has  real 
dimension four: two translations, one phase and the scale factor
(the instanton radius).  By $\N=(2,2)$
supersymmetry this implies  four fermion zero modes, 
which explicitly demonstrates that the axial U(1) symmetry is anomalous, and only a discrete subfactor of it survives,
namely,
 $$\U(1) \to \Z_4\,.$$ 
Further braking $\Z_4 \to \Z_2$ due to the bifermion condensate
is dynamical,  due to strong coupling.

For homogeneous spaces, such as $\mathbb{CP}^{1}$,  the choice of the base point for the homotopic cycle
 is  irrelevant. In field theory this is the point where the boundary at infinity maps 
 onto the target manifold. For the $\mathbb{CP}^{2}/{Z}_{2}$ orbifold we have to make a distinction 
 between two cases: (i) the base is the fixed point $(1,0,0)$; (ii) the base is any other point. 
In the case (ii)
 the  extra moduli space generated by the $\SU(2)$ symmetry between the coordinates $z_2$ and $z_3$ of 
 the orbifold moves the point at infinity, and thus does 
 not generate any additional zero modes in the instanton moduli space.
If the base is instead the Abelian fixed point (case (i)), the $\SU(2)$ symmetry 
generates zero modes. The total
number  of real bosonic zero modes for the instanton with
the  boundary at the fixed point is thus six.\footnote{
Alternatively, we could  establish this fact by considering instantons in $\mathbb{CP}^{2}$, 
and then reducing by   $\Z_2$. 
Instantons in $\mathbb{CP}^{2}$ have six bosonic 
zero modes -- the position, the size, the phase and two other 
extra coordinates that correspond to the choice of an $S^2$ inside $\mathbb{CP}^2$ --
and six fermion superpartners. If the base point is invariant under the orbifold projection, the six zero modes remain  in the orbifold, even if the metric is not exactly that of $\mathbb{CP}^{2}/{Z}_{2}$.}

We want to explicitly derive  the instantons solutions in the HT model.
At $m=0$, the isometry group of our sigma model is SU(2)$_{c+f}$,
 acting in the standard way on the three-sphere
parametrized by $(\alpha,\beta,\gamma)$. The coordinate $\omega$
does not transform under this SU(2).
The isometry group of $\mathbb{CP}^2$ 
with the standard metric is SU(3), which is much larger.

From the topological standpoint  the $\mathbb{CP}^2/{Z}_2$
instantons should  be rather similar to the  $\mathbb{CP}^2$ case.
The only difference is that in $\mathbb{CP}^2/{Z}_2$
configurations with the $\mathbb{CP}^2$
 topological charge $1/2$  
are allowed. 

In the sigma model under consideration the metric is very different
from that on the homogeneous  $\mathbb{CP}^2$ space. It
has much fewer
isometries. Hence, the explicit instanton solutions  are different.
Also, instantons, in principle, will change if we vary
  the vacuum expectation value of $\omega$. There is no symmetry of the
theory which relates two different values of $\omega$.
Let us consider some explicit instanton {\em ans\"atze}. In what follows $m=0$.

\subsubsection{Instanton A} 
\label{instA}

One possibility is to consider configurations at
$\omega=1$ (which corresponds to $\kappa=0$)
and generic $(\alpha,\beta)$. These are exactly the instantons
of the classical $\mathbb{CP}^1$ sigma model  at $\omega=1$.
Let us parametrize by $(\rho,\varphi)$ the two-dimensional world sheet.
We can use the {\em ansatz}
\beq 
\alpha(\rho,\varphi)=\alpha(\rho) \, , \qquad
\beta(\rho,\varphi)= \varphi \, .  
\label{dom7}
\eeq
Then the action is given by
\beq 
S=4 \pi \, r \,  \int \rho \, d\rho \left( (\partial_\rho \alpha)^2 +\frac{\sin^2 \alpha \cos^2 \alpha}{\rho^2} \right) \, .
\label{f64}
 \eeq
The Bogomol'nyi completion is
\beq  
S=
4 \pi \, r \,  \int d \rho \left[  \rho \,   \left( 
 \partial_\rho \alpha +\frac{\sin \alpha \cos \alpha}{\rho}  \right)^2 +  \partial_\rho ( \cos^2 \alpha )
 \right].
 \label{f65}
  \eeq
For this action the instanton solution is given by the well-known result
\beq 
\alpha=\frac{1}{2} \arccos \left(\frac{\rho^2-a^2}{\rho^2+a^2} \right)\, ,
\label{finss}
\eeq
where $a$ is the instanton size.
The action for this instanton is 
\beq
S_{\rm inst} = 4 \pi r\,.
\label{actins}
\eeq
It is easy to check that this configuration has at least four real bosonic zero modes:
the position, the size $a$ and a phase corresponding to a constant shift in $\beta$.
We will see that it can be interpreted as a composite instanton. 
Therefore, in fact  it must have more zero modes than those indicated above.
The situation is similar to the composite kink discussed in Sect.~\ref{kkki}.

\subsubsection{Instanton B} 
\label{instB}

Now, let us try  another simple {\em ansatz}.
Choose $\alpha=0$ and a nontrivial $(\kappa,\zeta)$,
\beq 
\kappa(\rho,\varphi)=\kappa(\rho) \, , \qquad
\zeta(\rho,\varphi)= \varphi \, .  
\label{fnontr}
\eeq
Then the action is given by
\beq
S=
 2 \pi r \,  \rho \,  \int d\rho  \, \left[ 4A (\partial_\rho \kappa)^2 +
\frac{1}{\rho^2} \frac{(1-\kappa^2)^2 (1-(1-\kappa^2)^4) }{2 -4  \kappa^4+4  \kappa^6 -\kappa^8}
\right] \, ,
\label{fnontrp}
 \eeq
and its Bogomol'nyi completion takes the form
 \beqn
 S
 &=&
 2 \pi r \,   \int d \rho \left[   
 \rho \,  \left( 2 \sqrt{A} (\partial_\rho \kappa)-\frac{1}{\rho} 
 \sqrt{ \frac{(1-\kappa^2)^2 (1-(1-\kappa^2)^4) }{2 -4  \kappa^4+4  \kappa^6 -\kappa^8}}
\,\,\, \right)^{\! 2}  \right. 
 \nonumber\\[4mm]
&-&
\left.   \partial_\rho \left( 
\kappa^2 \, (\kappa^2-2 )
\right) 
\rule{0mm}{8mm}
\right] \, .
\label{bogcoB}
\eeqn
 The  solution to the equation
 \beq
  2 \sqrt{A} (\partial_\rho \kappa)=\frac{1}{\rho} 
 \sqrt{ \frac{(1-\kappa^2)^2 (1-(1-\kappa^2)^4) }{2 -4  \kappa^4+4  \kappa^6 -\kappa^8}}
 \label{soltoe}
 \eeq
 can be found numerically.
The instanton action in this case is
 \beq
S_{\rm inst} = 2 \pi r\,.
\label{actinsp}
\eeq

 This instanton has a total of six bosonic zero modes:
 the position, the size and three extra zero modes which can
 be generated by using the SU(2)$_{c+f}$ rotation
 (one of these modes corresponds to a trivial constant shift in $\zeta$).
Therefore,  in the vacuum with $\omega=0$
 the dimension of the bosonic part of the instanton moduli space is six.

The instanton A is a configuration with the topological charge 
 twice larger than that   of  the instanton B. The instanton B is, therefore,
  the elementary instanton,
while the instanton A is a composite object.
The instanton A is not the most general instanton with
topological charge 2. It is just a very special solution which
can be found by a trivial embedding of the $\mathbb{CP}^1$ instanton.

\subsection{Transversal fluctuations}
\label{tf}

As was mentioned previously, 
fixing the position in the noncompact part of the manifold 
(the distance in the case of 2-strings), and then quantizing the 
compact part  is   an approximation. In quantum field theories in 
$2+1$ dimensions or higher this strategy is easily justifiable since distinct 
vacua labeled by different expectation values of scalar fields
obviously form separate nonoverlapping sectors in the Hilbert space.
 In $1+1$ 
dimensions the situation is subtler, and we must  check the effect of long-range 
transversal fluctuations. 
A free scalar field in $1+1$ dimensions has a correlation function 
\beq \langle \varphi(0) \varphi(z)\rangle \propto \log z \, . \eeq
 At large distance it diverges; therefore,  it seems impossible   to set
 $\varphi(z) $ to constant (equal to $\varphi_0$)
 at every point $z$. Translated in our context, this seemingly 
 implies that the string position 
 cannot be set constant on the   world sheet. 

To regularize the problem
one can consider
a flux tube with a {\em finite} length $R$, attached to
some probe infinitely massive  monopole and
antimonopole. The quantum mechanical wave
function of the flux tube connecting the probe charges has a nonvanishing
width $\tilde \ell $, which was computed in \cite{lmw},
\beq 
\tilde \ell^2  =\frac{1}{\pi T } \, \ln \frac{R}{\lambda}  \, , 
\label{tilel}
\eeq
where $T=2 \pi \xi$ is the flux tube tension and $\lambda$ is a parameter
which is of the same order of magnitude as
 the intrinsic thickness of the string $\ell \approx 1/(e_3 \sqrt{\xi})$,
beyond which the string model is no longer applicable to
the flux tube. 

The intrinsic string thickness $\ell$ is the parameter that must be 
compared with the width of transversal fluctuations. 
We thus obtain an estimate for the critical distance $R_c$ 
at which the transversal fluctuations become comparable with the intrinsic string thickness,
\beq 
R_c \approx  \frac{c}{e_3 \sqrt{\xi}}  \exp \left(\frac{1}{e_3^2}\right)
\approx  c \, \ell \, \exp \left(\frac{1}{e_3^2}\right)
\, ,
\label{dom9}
\eeq
where $c$ is a positive constant.
In the limit of the weak bulk coupling, $e_3^2\ll 1$, we have $R_c\gg \ell$. 
If the string length $R$ is smaller than $R_c$,
it is fully legitimate  to treat the component vortices
as coincident and to quantize just the compact part of the moduli space.

Note that $R_c$ is of the same order of magnitude
of $1/\Lambda_{1+1}$. This is the natural infrared cutoff
for the quantization of coincident vortices.
In the mass-deformed theory with $\dm \gg \Lambda_{1+1}$,  
it is possible to consider flux tubes that are short enough so
that the transversal fluctuations are completely irrelevant.

In the quantum case $\dm \rightarrow 0$ one must be more careful.
  Quantization of the internal manifold gives rise to states -- the kinks -- with thickness $1/\Lambda_{1+1}$
   and this is exactly the length scale where the transversal fluctuations are as 
   large as the string thickness. 
We can trust the result of the previous approximation (i.e. keeping fixed the distance and then quantize the internal manifold) only if the internal manifold does not vary considerably if the distance changes by an amount comparable with the string thickness.

\section{Renormalization group flow: an attempt}
\label{rg}

The renormalization group (RG) flow for nonlinear sigma models
with generic metric was studied in \cite{friedan,agfm}.
The basic idea is that the RG flow   changes  geometry of the sigma model.
In the homogeneous spaces case (such as $\mathbb{CP}^{N-1}$) the change 
of geometry amounts just to a change
in an overall factor in front of the metric.
This factor is identified as the coupling coupling constant; it describes the overall scale of the
target space. Say, for $\mathbb{CP}^1$ this is related to the radius of the sphere $S_2$.
For more general geometries all  elements of the metric $g_{ij}$, not just the overall scale,
change due to the RG flow.
The renormalization is governed by 
 a $\beta_{ij}$ function which
generalizes the well-known $\beta$ function  in the  homogeneous spaces,
\beq 
\mu \frac{\partial g_{ij}}{\partial \mu} =\beta_{ij} \, , \qquad \beta_{ij}=R_{ij} \, ,
\label{RG} 
\eeq
where $R_{ij}$ is the Ricci tensor 
\footnote{In the mathematical literature, 
this corresponds to the Ricci flow.
Ricci flow in relation to vortex moduli space has been considered
in a classical context in \cite{manton}.}. 
Equation (\ref{RG}) is valid at one loop.
The two loop contribution is non-zero and is proportional to \cite{agfm}:
\beq \mathcal{D}^k \mathcal{D}_k R_{ij} +2 R_{ikjl} R^{kl} + 2 R_{ik} R_j^k \, ,\eeq
where $\mathcal{D}_k$ is the covariant derivative
 from the standard Christoffel symbols
 obtained from the metric $g_{ij}$.

As usual in this paper, we put $z=0$, so that the metric and the Ricci tensor depend on four
coordinates.
Then the Ricci tensor and the metric tensor
have a similar structure which will allow us
to write the RG flow equations in a relatively simple form (\ref{a1}) -- (\ref{a3}).

The HT metric at z =0 is only topologically equivalent  to that of   $\mathbb{CP}^2/Z_2$, while
geometrically they are  different. That's why in $\mathbb{CP}^2$ the RG flow reduces 
to a variation of a single parameter, while in the HT case we
will have to introduce three functions.
In addition to the RG change  
of the overall scale factor (which certainly does take place),  
geometry gets ``distorted"
in all   directions too. The RG variations are faster
in some directions and slower in others.
If it were not for these distortions we will have to conclude 
that $r$ runs in the same way as in the $\mathbb{CP}^2$ model.

After these preliminary remarks we move on to 
consider a class of metrics which generalize the one  
obtained in Sect.~\ref{tkintt}, 
\beq 
 f_1(\kappa) d \kappa^2+
 f_2(\kappa) \left[ d \alpha^2 +\left(\frac{\sin 2 \alpha}{2}\right)^2 d \beta^2 \right]+
 f_3(\kappa)  \left(d \zeta -2 (\sin^2 \alpha) \, d \beta \right)^2 \, ,
 \label{mmetric}
 \eeq
 where $f_{1,2,3}$ are functions of $\kappa$.
Those functions that we found in Sect.~\ref{tkintt}  correspond to
\beqn
f_1&=&
r \frac{4 (\kappa^8-4 \kappa^6 +4 \kappa^4-2)}{\kappa^6 -4 \kappa^4 +6 \kappa^2 -4}  \, , 
\nonumber\\[2mm]
 f_2&=&
 r \, 2 (\kappa^2-1)^2 \, , 
\nonumber\\[2mm]
 f_3&=&
 r    \frac{\kappa^6 -4 \kappa^4 +6 \kappa^2 -4}{\kappa^8-4 \kappa^6 +4 \kappa^4-2}   (\kappa^2-1)^2  \kappa^2 \,,\label{bou}
\eeqn
with $0 \leq \kappa \leq 1$.
The metric of $\mathbb{CP}^2/\mathbb{Z}_2$
is, instead, given by 
\beq f_1=r \, , \qquad f_2=r \cos^2 \kappa \, ,
 \qquad f_3=r \frac{\sin^2(2 \kappa)}{16} \, ,\eeq
 with $0 \leq \kappa \leq \pi/2$.
\begin{figure}[h]
\begin{center}
$\begin{array}{c@{\hspace{.2in}}c@{\hspace{.2in}}c} \epsfxsize=1.5in
\epsffile{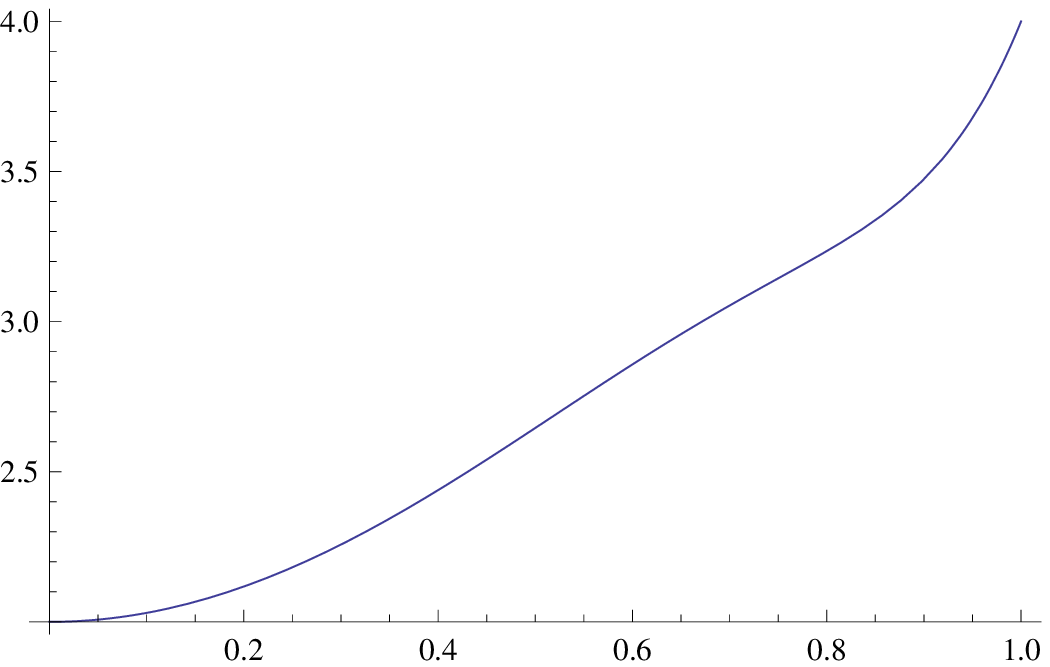} &
    \epsfxsize=1.5in
    \epsffile{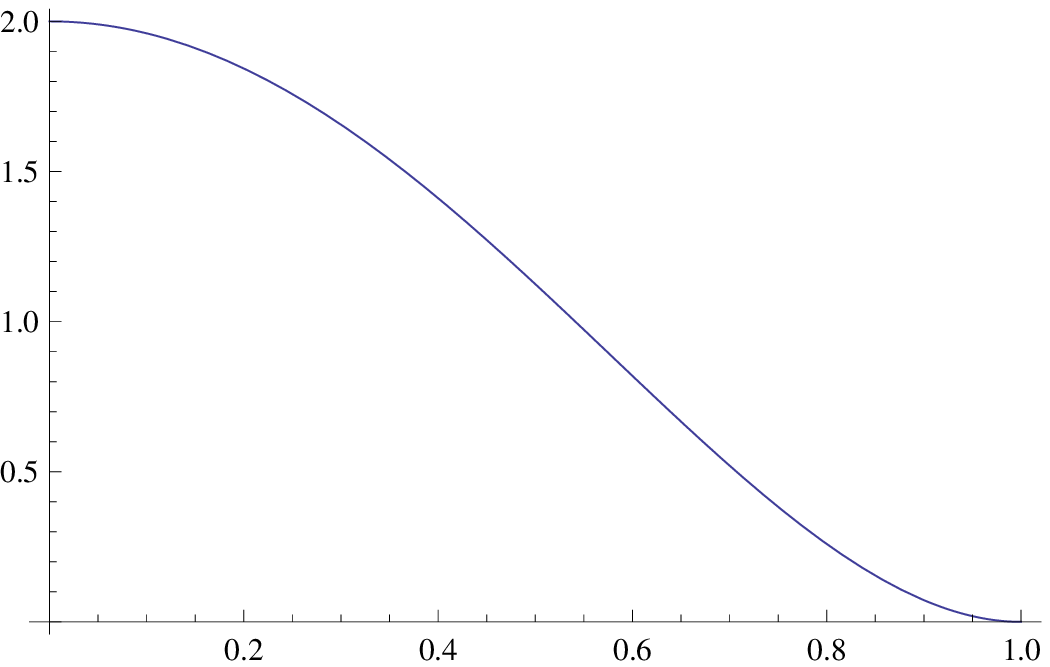} &
     \epsfxsize=1.5in
    \epsffile{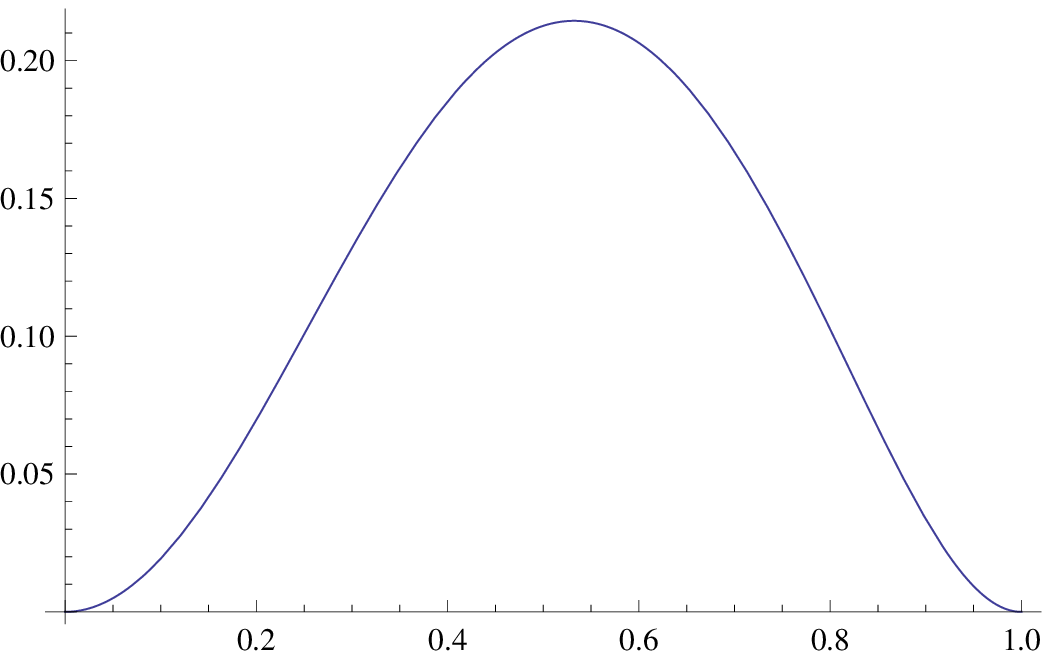}
\end{array}$
\end{center}
\caption{The functions  $f_1(\kappa),\,\, f_2(\kappa),\,\, f_3(\kappa)$ in Eq. (\ref{bou})
for $r=1$.} \label{profili}
\end{figure}

It is important to stress that in the metric (\ref{mmetric}) there is a freedom to redefine
the variable $\kappa$ by an arbitrary function.
In other words, the  above parametrization
in terms of   three functions $f_1$, $f_2$ and $f_3$ is redundant.
To fix this redundancy we can introduce a new variable,
\beq 
\lambda(\kappa)=\int_0^\kappa \sqrt{ f_1(\eta) }  d \eta \, , 
\label{nonred}
\eeq
and then express $f_2$ and $f_3$ in terms of $\lambda$.
The resulting metric can then be written as
\beq   d \lambda^2+
 f_2(\lambda) \left[ d \alpha^2 +\left(\frac{\sin 2 \alpha}{2}\right)^2 d \beta^2 \right]+
 f_3(\lambda)  \left(d \zeta -2 (\sin^2 \alpha) \, d \beta \right)^2 \, . 
 \label{mmetric2}
 \eeq
The functions $f_2(\lambda)$ and $f_3(\lambda)$, together with the range
of the
the  $\lambda$ variation,
 $$0<\lambda<\lambda_f\, ,$$
  specify the metric in a 
a way that is not redundant.

However,   to write the RG  equations,
it is inconvenient to fix the redundancy as in Eq.  (\ref{mmetric2}). 
A nice property of the class of metrics  (\ref{mmetric}) is that the
we can write the one-loop RG  equations as a system of differential equation for $f_{1,2,3}$.
If we compute the Ricci tensor from the metric (\ref{mmetric}) and  plug it back
in Eq. (\ref{RG}), we get the following system of equations:
\beqn 
&&
r \mu  \frac{\partial f_1}{\partial \mu} -\frac{f_3''}{2 f_3}+ \frac{(f_3')^2}{4 f_3^2}
+\frac{f_1' \, f_3'}{4 f_1 f_3} -\frac{f_2''}{f_2}   +\frac{(f_2')^2}{2 f_2^2}+ \frac{f_1' f_2'}{2 f_1 f_2} =0 \, , 
\label{a1}
\\[3mm]
&&
 r \mu  \frac{\partial f_2}{\partial \mu} -\frac{f_2''}{2 f_1}-\frac{f_2' f_3'}{4 f_1 f_3} + \frac{f_1' f_2'}{4 f_1^2}
-8 \frac{f_3}{f_2} +4 =0  \, ,   
\label{a2} 
\\[3mm]
&&
   r  \mu  \frac{\partial f_3}{\partial \mu} -\frac{f_3''}{2 f_1}+\frac{(f_3')^2}{4 f_1 f_3}-\frac{f_2' f_3'}{2 f_1 f_2}
+\frac{f_1'  f_3'}{4 f_1^2} +8 \frac{f_3^2}{f_2^2} =0 \, ,  
\label{a3}
\eeqn
where the prime denotes differentiation with respect to $\kappa$.
This is a nontrivial property for the metric of the form (\ref{mmetric});
usually the Ricci tensor is a very complicated expression in terms of the metric.
In our case it is quite simple, that's the reason why we managed to convert
Eq. (\ref{RG}) in 
(\ref{a1}) -- (\ref{a3}).

When we try to solve Eqs. (\ref{a1}) -- (\ref{a3}),
we find problems nearby $\kappa=1$, corresponding to the 
$(1,1)$ vortex. The solution for the profile $f_1$ is highly
unstable and is not trustworthy. 

\begin{figure}[h]
\begin{center}
\leavevmode
\epsfxsize 6 cm
\epsffile{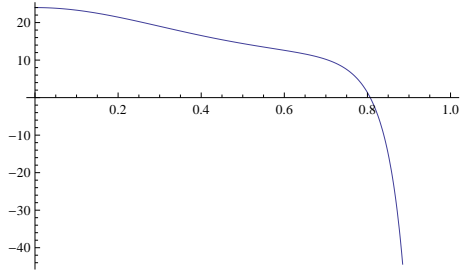}
\end{center}
\caption{\footnotesize Scalar curvature as a function of $\kappa$ for $r=1$.
At $\kappa=1$ the scalar curvature $R$ diverges.
This is a signal of a singularity associated with the $(1,1)$ vortex. }
\label{curvature}
\end{figure}

We would like  to emphasize that, strictly speaking, we can trust
Eq.~({\ref{RG}}) for the RG flow only far away from 
$\kappa=1$ (and, remember,  $\kappa=1$ corresponds to the $(1,1)$ string).
The problem is that the one-loop expression is trustworthy only
in the limit of small scalar curvature $R$,
\beqn 
R
&=&
\frac{1}{r} \left( 
\frac{8}{f_2} -\frac{8 f_3}{f_2^2} +\frac{f_1' f_2'}{f_1^2  f_2} +\frac{(f_2')^2}{2 f_1 f_2^2}
+\frac{f_1' f_3'}{2 f_1^2 f_3} \right.
\nonumber\\[3mm]
&-&
\left.
 \frac{f_2' f_3'}{f_1 f_2 f_3} +\frac{(f_3')^2}{2 f_1 f_3^2}
-\frac{2 f_2''}{f_1 f_2}-\frac{f_3''}{f_1  f_3}
\right).
\eeqn
In our example this quantity  diverges at $\kappa=1$ as shown in Fig.~\ref{curvature}.
Hence, we can not one-loop calculation in this domain. This is probably the origin
of the difficulties that we find when we try to solve
(\ref{a1}) -- (\ref{a3}) numerically.

This is also consistent with the fact that the subspace corresponding to
coincident vortices is not a manifold nearby the $(1,1)$ vortex
(there is a conical singularity already in the topology).
A possible way out is to consider the full metric, including
the $z$ dependence. It could be that the singularity in the metric
which makes the scalar curvature to diverge will disappear
once we consider the full six dimensional metric
and that this will make the RG flow calculation well defined
\footnote{It is important to stress that the moduli space of coincident
vortices has already a singularity in the topology in correspondence
of the $(1,1)$ vortex, because the space, strictly speaking,  is not  a manifold
in the neighborhood of this point. The singularity in the topology
disappears if we consider the full moduli space with arbitrary separation
and orientation \cite{recon}; the full moduli space is then topologically
a manifold in the neighborhood of every point.}. 

It is also possible that the divergence of the curvature nearby
the $(1,1)$ vortex signals a general problem in studying the physics
of that vacuum in a weakly coupled regime. 
A more detailed study of the full six-dimensional sigma
model would be desirable in order to understand this point.
In Appendix another section of the moduli
space is considered; it corresponds to antiparallel vortices at arbitrary distance $z$.
Also in this sub-manifold the curvature in correspondence
of the $(1,1)$ vortex is diverging.

\section{Conclusions}
\label{conclu}

We studied several aspects of coincident non-Abelian
vortex strings   using an effective description proposed 
in \cite{ht1,ht2},  suggested by the D-brane realization of $\mathcal{N}=2$ SQCD
in type II A string theory \cite{hw,witten97}. 
In the case of coincident strings we argued that the HT model   describes,
 in a consistent way,
a number of ``protected" aspects  of the world-sheet dynamics, such as the number of vacua, 
their symmetries and the masses of the  confined monopoles.

Topology of the string moduli space in
field theory and the one found from the brane
construction \cite{ht1,ht2} coincide \cite{knp,knp2,knp3}.
The situation with the metric is more murky;
we know that for large string separations the two metrics 
are different. For this reason the HT model cannot be viewed as fully realistic.
Despite this, we claim that the results presented in this paper 
would stay  valid   in the ``true'' model of multiple strings. The most important  of them is
the fact that
composite monopoles can be confined on composite strings, and retain their BPS nature.

The HT model emerges as a valuable (and in some instances, unique) tool in analyzing
non-Abelian strings. On the other hand this model is of a significant interest {\em per se}.
There are two obvious problems which should be addressed in the future:
large-$N$ solution of the HT model in the regimes (i) $k \sim N$ and (ii) $k\sim N^0$.

\section*{Acknowledgments}

We are grateful to D. Tong, A. Vainshtein, W. Vinci and A. Yung for very useful discussions.
 
The work of MS was supported in part by DOE grant DE-FG02-94ER408.

\section*{Appendix: Antiparallel-flux strings} 
 \renewcommand{\theequation}{A.\arabic{equation}}
\setcounter{equation}{0}

The general six-dimensional metric for 2-strings is difficult to write in
an explicit way. The main topic of this paper was the metric restricted to $z=0$,
 a much simpler task. There is another natural section of the moduli space where it
is easy to write down the metric and the potential; it 
can be obtained restricting to  $\omega=0$. It
corresponds to elementary vortices with the opposite internal orientations, i.e. the composite system of
(1,0) +(0,1).
 
The following gauge fixing can be used:
\beqn
 a_i
 &=& r^{1/2}
 \, (\cos \alpha, e^{i \beta} \sin \alpha) \, ,
 \nonumber\\[3mm]
b_i
&=& r^{1/2}
 \, ( e^{-i \beta} \sin \alpha, - \cos \alpha) \, , \qquad
 Z= \left(\begin{array}{cc}
z  & 0  \\[1mm]
0  & -z  \\
\end{array}\right) \, .
\eeqn
By a straightforward calculation similar to those  in
Sects. \ref{tkintt} and \ref{ttmtt},  we can find both the metric and  potential
for this section.
The kinetic term is
\beq 
2 (\partial_\mu z)^2+8 r \, \frac{z^2}{r^2+4 z^2} \, 
 \left[(\partial_\mu \alpha)^2 +\left(\frac{\sin 2 \alpha}{2}\right)^2 (\partial_\mu \beta)^2 \right] \, ,
 \label{pa2}
\eeq
while the potential induced by the twisted mass term is
\beq 
V=   8 m^2 r \left(\sin^2 2 \alpha\right)\,  \frac{z^2}{r^2+ 4 z^2} \, .
\eeq
From these expressions it is easy to infer that the kinetic term for the $S^2$ part 
approaches the asymptotic value in a power-like manner,
instead of the exponential law  we would expect in the gapped
bulk theory (this is a bad feature of the model).
We can also check that the interactions between the component strings
start to be relevant at $z\approx \sqrt{r}$, which is consistent with
the expected vortex thickness in the weakly coupled limit (this is a good feature).
  \begin{figure}[h]
\begin{center}
\leavevmode
\epsfxsize 6 cm
\epsffile{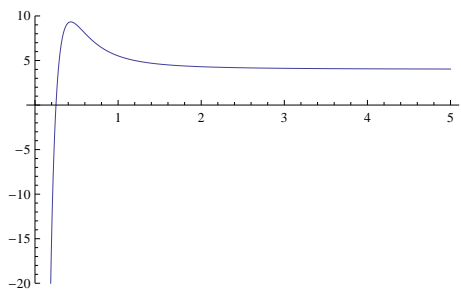}
\end{center}
\caption{\footnotesize Scalar curvature for the moduli space section
corresponding to the strings with the opposite values of $\vec{B}^3$,  
as a function of $z$ for $r=1$. }
\label{curvature2}
\end{figure}
 It is instructive to compute the scalar curvature for the metric (\ref{pa2}); we get 
\beq
 R= \frac{-2 r^3+28 r^2 z^2+48 r z^4+64 z^6}{r z^2 \left(r+4 z^2\right)^2} 
 \rule{0mm}{9mm} \, .
 \label{pa4}
 \eeq
This expression is plotted in Fig.~\ref{curvature2}. It diverges, $R\to -\infty$, 
at the point $z\to 0$. It is unclear what would happen if we could lift the restriction
$\omega =0$. 
In the full moduli space the scalar curvature at $z\to 0$ could still be finite, or
tend to $-\infty$ as in (\ref{pa4}).

\end{document}